\documentclass[12pt,preprint]{aastex}
\usepackage{psfig}

\shorttitle{}
\shortauthors{Dannerbauer et al.}

\begin{document}

\title{The faint counterparts of MAMBO mm sources near the NTT Deep Field}

\altaffiltext{1}{Based on observations collected at ESO (66.A-0268,
67.A-0249, 69.A-0539, 70.A-0518, 71.A-0584) at the VLA, and on observations
carried out with the IRAM PdBI.  IRAM is supported by INSU/CNRS (France),
MPG (Germany) and IGN (Spain).  The Very Large Array is a facility of
the National Radio Astronomy Observatory, which is operated by Associated
Universities Inc., under cooperative agreement with the National Science
Foundation.}

\author{H.~Dannerbauer\altaffilmark{2}, M.~D.~Lehnert\altaffilmark{2},
D.~Lutz\altaffilmark{2}, L.~Tacconi\altaffilmark{2},
F.~Bertoldi\altaffilmark{3}, C.~Carilli\altaffilmark{4},
R.~Genzel\altaffilmark{2,5} \& K.~M.~Menten\altaffilmark{3}}

\altaffiltext{2}{Max-Planck-Institut f\"ur extraterrestrische Physik,
Postfach 1312, D-85741 Garching, Germany}

\altaffiltext{3}{Max-Planck-Institut f\"ur Radioastronomie, Auf dem
H\"ugel 69, 53121 Bonn, Germany}

\altaffiltext{4}{National Radio Astronomy Observatory, P.O. Box O,
Socorro, NM 87801, U.S.A.}

\altaffiltext{5}{Department of Physics, Le Conte Hall, University of 
California, Berkeley, CA 94720, U.S.A.}

\begin{abstract}

We discuss identifications for 18 sources from our MAMBO 1.2mm survey of
the region surrounding the NTT Deep Field. We have obtained accurate
positions from Very Large Array 1.4\,GHz interferometry and in a
few cases IRAM mm interferometry, and have also made deep BVRIzJK
imaging at ESO.  We find thirteen 1.2mm sources associated with
optical/near-infrared objects in the magnitude range K=19.0 to 22.5,
while five are blank fields at K$>$22.  We argue from a comparison
of optical/near-infrared photometric redshifts and radio/mm redshift
estimates that two of the thirteen optical/near-infrared objects
are likely foreground objects distinct from the dust sources, one
of them possibly lensing the mm source.  The median redshift of the
radio-identified mm sources is $\sim$2.6 from the radio/mm estimator, and
the median optical/near-infrared photometric redshifts for the objects
with counterparts $\sim$2.1.  This suggests that those radio-identified
mm sources without optical/near-infrared counterparts tend to lie at
higher redshifts than those with optical/near-infrared counterparts.
Compared to published identifications of objects from 850$\mu$m surveys
of similar depth, the median K and I magnitudes of our counterparts are
roughly two magnitudes fainter and the dispersion of I-K colors is less.
Real differences in the median redshifts, residual mis-identifications
with bright objects, cosmic variance, and small number statistics are
likely to contribute to this significant difference, which also affects
redshift measurement strategies. Some of the counterparts are red in J-K
($\gtrsim$20\%), but the contribution of such mm objects to the recently
studied population of near-infrared selected (J$_s$-K$_s$$>$2.3) high
redshift galaxies is only of the order a few percent. The recovery rate
of MAMBO sources by pre-selection of optically faint radio sources is
relatively low ($\sim$25\%), in contrast to some claims of a higher rate
for SCUBA sources ($\sim$70\%).  In addition to this difference, the
MAMBO sources also appear significantly fainter ($\sim$1.5 magnitudes
in the I-band) than radio pre-selected SCUBA sources.  We discuss
basic properties of the near-infrared/(sub)mm/radio spectral energy
distributions of our galaxies and of interferometrically identified submm
sources from the literature. From a comparison with submm objects with
CO-confirmed spectroscopic redshifts we argue that roughly two thirds
of the (sub)mm galaxies are at z$\gtrsim$2.5. This fraction is probably
larger when including sources without radio counterparts.

\end{abstract}

\keywords{galaxies: formation --- galaxies: high-redshift --- galaxies: starburst --- infrared: galaxies --- submillimeter}

\section{Introduction}
\label{Introduction}

With the discovery of distant submm and mm galaxies in blank field
surveys and cluster lens assisted surveys \citep[see][for a review and
references]{blain02}, it was immediately recognized that this population
holds important clues for the understanding of the formation and
evolution of galaxies. A significant part of the cosmic submm background
is produced by these objects that must be extremely luminous (L$_{IR}\sim
10^{12-13}$L$_\odot$) distant (z$\gtrsim$1) infrared galaxies powered
by intense star formation and/or powerful AGN. X-ray data argue in most
cases against (Compton-thin) AGN and in favor of intense star formation
dominating their luminosity \citep[e.g.,][]{alexander03}. These high star
formation rates ($\approx$100-1000\,M$_\odot$/yr) and the similarity of
co-moving space densities of submm sources and local ellipticals suggest
that they are likely indicating the formation of massive spheroids. This
opens a direct route to locating the formation of spheroids between
the two extremes of an early formation similar to the classical
`monolithic collapse', and a late formation in hierarchical merging. More
specifically, properties and mass functions for these high redshift
objects are robust tests for current hierarchical models of galaxy
formation \citep{guiderdoni98,kauffmann99,somerville01,baugh03}. Indeed
there is evidence \citep[e.g.,][]{genzel03, neri03} that these models
need some modification to reproduce the space densities of the massive
high redshift submm galaxies and their quiescent phases that must exist
since their star formation rates and gas content suggest a duty cycle
of the bright (sub)mm phase well below one.

Much of this promise will only fully come to fruition after a difficult 
process
of identification and spectroscopic redshift determination. Based on
optical spectroscopy, significant progress in the identification work
was done by \citet{chapman03a} who presented 10 redshifts of radio
identified counterparts of SCUBA galaxies.  But still, less than
a dozen optical/near-infrared redshifts for suggested counterparts
have been confirmed by CO observations as the true redshift of the
submm source \citep[][Greve et al. in prep, and SMM02399-0314, 
Kneib in prep.]{frayer98,
frayer99, neri03}.  Accurate positions from radio or mm interferometry
are available for just a few dozen non-radio-preselected submm
galaxies \citep[e.g.,][]{downes99, smail00, eales00, gear00,
lutz01, ivison02, ledlow02, webb03a,webb03b} and very few mm-selected galaxies
\citep{bertoldi00, dannerb02}.  These accurate positions are indispensable
for reliable optical/near-infrared identifications since several
possible optical/near-infrared counterparts are usually found in the
several arc second radius error circles of the (sub)mm surveys. This
identification step is the subject of the present work. Of the two
interferometric identification methods in use for (sub)mm sources, mm
interferometry has the advantage of directly and unambiguously locating
the dust emission, but it is very time consuming with current instruments,
with tens of hours typically needed for a single object in the small
field of view.  Because of the tight radio/far-infrared relation for
star-forming galaxies \citep[e.g.,][]{dejong85, helou85, condon92}, radio
interferometry is the next best option, with the advantage of the large
VLA primary beam covering one of the currently typical (sub)mm survey
fields entirely. Of the brighter (sub)mm sources from present surveys,
deep 1.4GHz VLA maps will detect all but the highest redshift ones
\citep{carilli99,carilli00a,barger00}, and the risk of false associations
of (sub)mm and radio sources is modest for the 1.4GHz source counts at
the relevant flux levels of tens of $\mu$Jy \citep[e.g.,][]{richards00}.

We are building on our mm survey (Bertoldi et al. 2003, in preparation)
that uses the MAMBO array \citep{kreysa98} at the IRAM 30m telescope to
cover three fields, the Lockman Hole, Abell 2125, and a region centered
on, but larger than the NTT Deep Field \citep[NDF,][]{arnouts99}. This
paper focuses on the NDF region.  \citet[hereafter Paper~I]{dannerb02}
have presented results for three of the brightest NDF mm sources using
the IRAM Plateau de Bure mm interferometer (PdBI) and BVRIK imaging.
We now increase to 18 the number of NDF mm sources with accurate
positions by using our VLA interferometry, and introduce additional
z- and J-band optical/near-infrared imaging obtained since publishing
Paper~I. We thus identify counterparts and start assembling optical to
radio spectral energy distributions (SEDs) of the objects constraining
their nature and redshift.  The improved statistics is the basis for
a discussion of the properties of mm galaxies and their relation to,
or differences from, the submm population. Throughout the paper we
adopt the cosmological parameters $H_0$=70\,km\,s$^{-1}$\,Mpc$^{-1}$,
$\Omega_M$=0.3, $\Omega_\Lambda$=0.7.

\section{Observations, Data Reduction and Results}
\label{sect:obsandred}

The radio data play a key role in the identification presented in
this paper. The NTT
deep field was observed at 1.4 GHz with the Very Large Array (VLA) in
April and May, 2001, for 15 hours in the B configuration (10 km maximum
baseline). Standard wide field imaging mode was employed (50 MHz total
bandwidth with two polarizations and 16 spectral channels).  The source
3C 286 was used for absolute gain calibration and 1224+035 was used for
phase and bandpass calibration.  The data were also self-calibrated
using sources in the NTT deep field itself.  Images were synthesized
and deconvolved using the wide field imaging capabilities in the AIPS
task IMAGR, and the primary beam correction was applied to the final
image. The full width at half maximum (FWHM) of the Gaussian CLEAN beam
was $7\arcsec \times 5\arcsec$ with a PA $= 0\arcdeg$. The rms noise
on the final image is between 13 and 15 $\mu$Jy over the 15\arcmin\
$\times$15\arcmin\ region covered by MAMBO.  The variation of the rms
noise of the map was not only caused by the usual beam attenuation and UV
plane coverage, but was also due to residual calibration errors because
of a relatively large number of bright sources (2 sources of $\gtrsim$
60 mJy/beam) and one spatial grouping of fainter but still relatively
bright sources to the northeast of the center of the map.  These residual
calibration errors manifested themselves as uneven noise in a series
of stripes in the final map.  About 1/3 of area of the final VLA map
was affected by this striping.  We describe in \S~\ref{sect:assoc}
how this was dealt with in identifying radio counterparts to the MAMBO
1.2mm detections.

We use the BVRIK imaging of a field centered on the NDF that is described
in Paper~I.  We have obtained additional z and J-band imaging in spring
2003. The z-band observations were carried out in `service mode'
on the nights of April 6 and 7, 2003 using the imaging spectrograph
FORS2 on UT4 of the VLT.  The data were taken as a sequence of dithered
exposures with a net integration time of 1 hour each covering 4 separate
pointings. The pointings were chosen such that a mosaic of the images
taken at the 4 pointings covers the whole of the K-band field centered
on the NDF.  The images were processed in the standard way but were
flat-fielded using images generated by masking out all objects with
surface brightnesses outside 1 sigma of the background noise and
then combined by taking the pixel median of all the frames without
the images being aligned.  This flat-field frame was then normalized
to one and then divided into each z-band image.  Conditions were
photometric throughout the observations and the seeing was typically
about 0.5 arc seconds.  The final calibration was determined through
observations of the spectrophotometric standard LTT7987 \citep{hamuy94}
for the z-band filter. However, given the very sensitive red response
of the FORS2 array, the z-band filter was calibrated by integrating the
spectral energy distribution of LTT7987 \citep{hamuy94} and of BD+17 4708
\citep{oke83} which is the absolute flux calibrator for the Gunn-z filter
\citep{schneider83}.  In making these estimates, both the response of the
CCD and the filter transmission (both of which are available on the ESO
web pages) were considered.  The average noise across the frame is such
that the 3$\sigma$ detection limits in a 2\arcsec\ diameter aperture
is z$_{BD+17 4708}$=26.38.  The absolute flux distribution of LTT7987
provides a direct calibration of the images in the AB-magnitude system
which yields z$_{AB}$=z$_{BD+17 4708}$-0.30.  J-band imaging was obtained
with SOFI at the ESO NTT between March 14 and 16 2003, covering the same
$13\arcmin\times 13\arcmin$ region as our K$_s$ image of the NDF to a
3$\sigma$ limiting magnitude for a 2\arcsec\ aperture of 24.0 in J on
the Vega system. Data reduction followed similar steps as described in
Paper~I, with the photometric zero point in J based on 2MASS. We used
common sources to bring these new data to the same astrometric system
as described in Paper~I, including the small correction described in
Paper~I from the USNO-A1 based optical/near-infrared astrometry of our
optical/near-infrared data to the radio reference frame.  All positions
quoted here include this correction and are on the radio frame.

\subsection{Association of MAMBO sources with radio sources}
\label{sect:assoc}

In order to obtain more accurate positions for the MAMBO sources
through use of their radio continuum emission, we have searched the
1.4\,GHz VLA data for peaks of at least 40\,$\mu$Jy ($\sim 3\sigma$)
in the region where the MAMBO sources are located, out to radii of
about 10\arcmin\ from the phase center.  Table~\ref{tab:assoc} lists
the MAMBO sources in the catalog as of early 2002 which have radio
sources with peak positions less than 7\arcsec\ from the nominal MAMBO
map position. This radius is more than half the FWHM of the MAMBO
beam and will not exclude real associations even in case of slight
systematic offsets. The table, like all our subsequent analysis, does
not include the well-detected quasar BR1202-0725 which was driving the
original selection of the public ESO NDF, and thus cannot be considered
an unbiased representative of the mm galaxy population. A more complete
description of the MAMBO data analysis will be given in Bertoldi et al.
(2003, in preparation). At $5\arcsec\times 7\arcsec$ resolution, our
B configuration data do not put interesting constraints on the extent
of the radio emission of the detected galaxies.  Table~\ref{tab:assoc}
also lists the corrected Poissonian probability that an association is a
chance coincidence, derived using the approach of \citet{downes86} which
corrects the simple Poissonian probability of an observed association
for the possibility of associations of different nature but similar
probability. In deriving these probabilities we adopt the raw counts
of $\geq$40$\mu$Jy peaks in the region of our VLA image covered by the
MAMBO data, which are significantly above the true 1.4GHz source counts
for a flux close to the detection limit. Using true radio source counts
\citep[e.g.,][]{richards00} would thus have underestimated the chance
of a false association. The probabilities listed together with the fact
that 42 sources from the original MAMBO list were searched for nearby
radio peaks suggest that several of the associations may be chance. This
statistical analysis, however, does not adequately consider that the
excess of $\sim$40$\mu$Jy peaks boosting the rate of chance coincidences
is largely due to peaks in the VLA map that cluster in certain regions,
due to uneven noise and striping because of residual calibration
errors. Considering this enables us to identify those associations that
are more likely to be spurious by inspection of the distribution of
$\geq$40$\mu$Jy peaks in the radio map.  The quality of an associated
radio source is indicated in Table~\ref{tab:assoc}, with `uncertain'
assigned to sources with an excess of similar brightness peaks near to
the suggested radio source, and in particular with alignments of peaks
due to striping. We also consider as uncertain radio peaks that do not
remain above 40$\mu$Jy in a radio image where larger scale variations are
partially removed by subtracting a 20\arcsec\ median smoothed version
of the radio image, and two faint radio peaks near sources No. 10 and
25 that are only above 40$\mu$Jy in this image to which we refer as the
'background-subtracted' image below. Of the identifications we consider
`good' below, all but source 13 are either above 4$\sigma$ in the original
radio map, or confirmed by PdBI mm interferometry.

Figure~\ref{fig:assoc} displays the mm--radio offsets corresponding to
the associations listed in Table~\ref{tab:assoc}. The dominance of real
associations is clear. There were no indications for systematic offsets.
The separations found are consistent with a typical MAMBO position
error of about 3\arcsec\/, and confirm the radius of 7\arcsec\ as a
sensible choice for searching for associated radio sources.  Overall,
interferometric positions are available for 18 out of 42 sources
in the NDF MAMBO map.  We consider 11 of these 18 reliable, either
through `good' radio peaks or through direct PdBI mm interferometry
(Paper~I). Interestingly, the PdBI data confirm the reality of two of
the more uncertain radio sources, increasing confidence in the rest of
these objects. The remaining MAMBO sources without radio associations
could be at high redshift (with corresponding radio fluxes below our
detection limit), could be spurious mm sources or have MAMBO fluxes
systematically overestimated because of the bias induced by noise in a
population with steep number counts. Given the signal-to-noise ratios
of the MAMBO detections all these explanations may contribute. For the
purpose of this identification paper, we do not discuss further objects
without an interferometric position. This induces a bias towards lower
redshift and brighter objects.

\subsection{Optical and near-infrared objects near the radio positions}

Images of $25\arcsec\times 25\arcsec$ size in BRzK for the MAMBO sources
with interferometric position are shown in Figure~\ref{fig:idplot1}.
Table~\ref{tab:idtable} lists the positions and BVRIzJK photometric data
of optical/near-infrared objects that are close to the interferometric
positions. The photometry is in 2\arcsec\ diameter apertures centered
on the positions which are derived from a first step SExtractor
\citep{bertin96} analysis of a noise-scaled co-addition of the
BVRIzJK images. Wavelength-dependent morphology can lead to small
offsets of the peaks in certain bands from these average positions
(Fig.~\ref{fig:idplot1}).  We have verified the reality of objects
in Table~\ref{tab:idtable} that are fainter than the 10$\sigma$
completeness limit and close to the detection limit by visual inspection
of the images in the various bands. We have retained objects that are
reliably detected individually in either B,z, or K, these images being
particularly deep and including the extreme wavelengths first detecting
blue or very red objects. For the measurements in the individual bands,
we indicate in Table~\ref{tab:idtable} cases where we consider the
SExtractor aperture magnitudes more uncertain, since the object is only
tentatively (indicated by `$\approx$') or not (indicated by `!') picked
up by the eye in the visual inspection.  For simplicity, we designate
the optical/near-infrared objects with a combination of the short MAMBO
number (Table~\ref{tab:assoc}) and a letter increasing with distance
from the interferometric position.  These letters are also indicated in
Fig.~\ref{fig:idplot1}.  The short MAMBO numbers used in this paper are
consistent with the NTT-MMnn numbers used by \cite{eales03} in their
comparison of 850$\mu$m and 1.2mm fluxes of MAMBO sources. BzK color
composites of the objects are shown in Fig.~\ref{fig:colorcomp}.

We have used the publicly available {\it hyperz} photometric redshift code
\citep{bolzonella00} to estimate redshifts from our seven band photometry.
We have verified the {\it hyperz} results for our specific photometric
dataset by comparison of photometric  redshifts to spectroscopic redshifts
for 63 objects between z$_{spec}$=0.07 and 1.69 in the NTT Deep Field and
its surroundings, for which spectroscopic redshifts are available from
our own projects in that region and from ESO Science Verification data
(http://www.eso.org/science/ut2sv/NDFs\_release.html). The dispersion
$\sigma$$_{zspec-zphot}$ of 0.17 is satisfactory, with few gross outliers.
No trend between the accuracy of our photometric redshift and the color
(R-K) is seen. We note however that {\it hyperz} does not contain an
observed SED of a ultra-luminous infrared galaxy (ULIRG; which the SED
of a mm sources may be similar), although it does allow for a range of
additional extinction to be added to the model which may approximate
a SED similar to that seen in ULIRGs. We notice a systematic effect
at z$\lesssim$1, in the sense that photometric redshifts tend to be
overestimated for sources with z$\lesssim$0.3 and underestimated in
objects with 0.7$\lesssim$z$\lesssim$1.0. This is similar to what is
seen in Subaru Deep Field photometric redshifts obtained with {\em
hyperz} \citep[][their Fig.~1]{kashikawa03}. These systematic effects
as well as the dispersion of z$_{spec}$-z$_{phot}$ suggest that for
individual bright objects the 1$\sigma$ errors from {\it hyperz} in
Table~\ref{tab:idtable} underestimate the real errors, but at a level
where the photometric redshifts are still valuable constraints for the
suggested counterparts and for nearby objects.

\subsection{Identification and notes on individual objects}

This section gives a case-by-case discussion of the individual
interferometrically located MAMBO sources and identifies the suggested
optical/near-infrared counterparts. We consider as possible counterparts
optical/NIR objects that are less than 2 times the interferometric
positional error (Table~\ref{tab:assoc}) away from the interferometric
position. Again, we have derived the corrected Poissonian probabilities
for a chance coincidence computed according to \citet{downes86} and
considering the search radius of twice the radio positional error.
Table~\ref{tab:idtable} lists these probabilities for the objects
inside the 2$\sigma$ search circle, adopting the K-band number counts
of \citet{totani01b}. Deriving probabilities this way ignores color
information and thus implicitly assumes a similar color distribution
of counterparts and field population.  In reality the colors differ
(\S~\ref{sect:optnir}), i.e. we are overestimating the probability
of a particular object being a random coinciding member of the field
population.

For convenience, salient properties of the counterparts are collected
in Table~\ref{tab:counter}, full color information can be extracted
from Table~\ref{tab:assoc}. Redshifts from the radio/submm spectral
index are based on the flux densities from Table~\ref{tab:assoc} and
the relation of \citet{carilli00b}. These redshifts are subject to the 
applicability of the far-infrared SEDs used to calibrate this relation, 
and may be overestimated if cool luminous infrared galaxies 
\citep[e.g.][]{chapman02b} are prominent in the (sub)mm population. The 
possibility of 'cool' galaxies is also intimately related to the role of 
lensing  in the identification of optical/near-infrared counterparts of 
submm sources \citep{blain99, chapman02c} -- as long as the redshift of 
the dust 
source is not confirmed through CO lines, certain configurations could
correspond either to a `cool' low z dust source or a `warm' high z dust 
source with a superposed lensing foreground object.
Therefore, for several objects, we present arguments
on the potential role of lensing by individual foreground galaxies.  These
are simple plausibility arguments based on the observed K magnitudes and
photometric redshifts, using after K-correction the K-band Faber-Jackson
relation of \cite{pahre98} to estimate the velocity dispersion and the
isothermal sphere approximation \cite[e.g.,][his eq. 4-14]{peacock99}
to estimate the radius of the Einstein ring.

\paragraph{MMJ120507-0748.1 (No. 03, g(ood) radio source)} The faint,
red counterpart 03a coincides with the 88$\mu$Jy radio source. A $\sim
1\arcsec$ NW/SE double peak morphology for 3a may be indicated in the
K$_s$ image but is not certain.  The optical/near-IR photometric redshift
z$\sim$2.1 is consistent with the redshift estimate based on the radio-mm
spectral index (z$_{CY}\sim$2.4). The higher estimate z=4.25 from the
submm/mm ratio \citep{eales03} is uncertain and consistent with these
lower values.

\paragraph{MMJ120508-0743.1 (No. 26, u(ncertain) radio source)} A
faint, uncertain radio source lies 4.6\arcsec\ from the nominal MAMBO
position. No optical/near-infrared counterpart coincides with this
radio source.

\paragraph{MMJ120509-0740.0 (34, u)} No optical/near-infrared counterpart
is detected at the position of the faint, uncertain VLA counterpart. In
the vicinity, an agglomeration of bright, normal galaxies is seen. The
photometric redshifts of these objects (z$_{phot}\approx$0.4-0.5) are
consistent with these galaxies perhaps being part of a group. At distances
of $>$4.8\arcsec\ from the interferometric position, lensing by these
individual galaxies is unlikely. Similarly, a small group as such is not
likely to produce a strong lensing effect \citep[e.g.,][]{hoekstra01}.
\citet{eales03} fail to significantly detect at 850$\mu$m this object
which is a 5$\sigma$ 1.2mm source in the MAMBO map and was seen at
similar flux (but only 2.2$\sigma$) in an independent short MAMBO on-off
observation \citep{eales03}. The very low 850$\mu$m/1.2mm flux ratio
as well as the non-detection of an optical/near-infrared counterpart may
suggest a very high redshift.

\paragraph{MMJ120510-0747.0 (10, u)} Near the nominal bolometer position
lies a small separation (1.5\arcsec\/) galaxy pair of a fairly compact
blue object (10a1) and a red (10a2) object.  The interferometric position
of the uncertain VLA source (seen only in the background-subtracted radio
image) is about 2.4\arcsec\ (1.9$\sigma$) north of 10a2 and slightly
more distant from 10a1.  It is thus marginally possible that one of the
objects of this pair is the counterpart of the MAMBO source. Photometric
redshifts suggest the blue object to be in the foreground rather than
being physically associated with the red object, there are no secondary
solutions for the photometric redshift which would put both at the same 
redshift. We tentatively identify
the source with object 10a2. The high photometric redshift, consistent
with the radio/mm estimate, is more uncertain than its nominal error given
the faintness and the need to decompose the two components of 10a. At the
observed K-band magnitude and estimated photometric redshift of 10a1,
the estimated Einstein ring radius is only about 0.30\arcsec\/, making
strong lensing of 10a2 at 1.5\arcsec\ very unlikely. The same is true
for other plausible redshifts. We note that 10a1 appears fairly compact
and blue, spectroscopic observations to look for evidence of an AGN may
be worthwhile.  In case of a strong AGN its current photometric redshift
estimate would be uncertain.

\paragraph{MMJ120516-0739.4 (36, g)} This object has the strongest radio
counterpart of a NDF 1.2mm source which coincides with a relatively
bright and extremely red (R-K$>$6.0) source.  The discrepancy between
the redshift estimate from the radio/mm spectral index ($\sim$0.8) and
the photometric redshift ($\sim$1.7) may suggest that the radio emission
is boosted by an AGN.

\paragraph{MMJ120517-0743.1 (25, u)} A faint radio counterpart is
not seen in the original VLA image but in the background-subtracted
version. It coincides with the position determined by mm interferometry
\citep[][see their note added in proof] {dannerb02}. Recently, a
faint K-band counterpart (22.5 magnitude) at the PdBI/VLA position was
detected in ultra-deep imaging performed with ISAAC at the VLT (Lehnert
et al. 2004, in preparation).  Similarly, the new z band image indicates
an extended source at the PdBI position with a maximum or companion to
the southwest. In B, V and R very faint extended emission is suggested at
the same position but only deeper imaging can reveal the reality of these
structures. The photometric redshift is 2.91 in good agreement with the
radio/mm redshift estimate and the mm/submm estimate of \citet{eales03}.

\paragraph{MMJ120519-0749.5 (01, g)}
Among the `blank field' sources not showing an optical/NIR counterpart
within 2$\sigma$ of the interferometric position, this is the blank
field object with the strongest radio counterpart, that is also a secure
radio source (72$\mu$Jy).  \citet{eales03} report a 4.3$\sigma$ 850$\mu$m
detection of this object, and infer a low 850$\mu$m/1.2mm flux ratio. This
suggests a very high redshift which under the SED assumptions made is
marginally consistent with the radio/mm redshift estimate.  The objects
a,d,e (Fig.~\ref{fig:idplot1}) marginally detected by SExtractor seem
to be due to noise or structure in the extended emission from the large
foreground object and were not included in Table~\ref{tab:idtable}.

\paragraph{MMJ120520-0738.9 (39, g)}
The radio counterpart is the second strongest in our sample and coinciding
to an ERO (R-K$=$5.2). In the K-band, it shows weak NW/SE extensions, with
possible variations at shorter wavelengths. The high quality photometric
redshift (1.32) and the radio/mm estimate agree well putting the object
towards the low end of the redshift distribution of our mm sources. Object
39f, 7.6\arcsec\ to the NW, has a very similar photometric redshift.

\paragraph{MMJ120522-0745.1 (18, g)}
The strong radio source which is 6\arcsec\/ SW from the nominal bolometer
position coincides with an elongated moderate photometric redshift ERO
(R-K$=$5.3), we here adopt it as counterpart of the MAMBO source noting
the relatively large positional offset.

\paragraph{MMJ120524-0747.3 (08, u)}
Two faint and uncertain radio sources lie within the MAMBO beam. Both
are blank fields in the optical/near-infrared. We adopt the $\mu$Jy
radio source closer to the nominal MAMBO position but consider the
chance of a misidentification to be significant, both due to the radio sources
being uncertain and the difficulty of deciding among them 
\citep[cf. also the case of SMM09431+4700]{neri03}.

\paragraph{MMJ120526-0746.6 (13, g)}
A faint radio source from a good region of the VLA image coincides with
an ERO (R-K$=$5.4). The K morphology appears distorted/curved. Both
photometric redshift and radio/mm spectral index put the source at
z$\approx$2.5.  This counterpart has very large J-K (and z-K) color,
we will discuss this below in \S~\ref{sect:optnir}.

\paragraph{MMJ120530-0741.6 (29, g)}
The interferometric position of the strong 85$\mu$Jy radio source
is in a group of three objects. Object 29a coincident with the
radio position is at I-K$=$3.3 a very red object in the definition
of \citet{ivison02}. 1.5\arcsec\/ to the south-west there is a bluer
source 29b, and 1.5\arcsec\/ to the north-west from 29a the extremely red
object 29c (I-K$>$4.3). The photometric redshifts of the three objects
are consistent with all being at z$\sim$2.4, in reasonable agreement
with the radio/mm estimate of $\sim$1.8. The three objects could be
physically associated, with separations (for a redshift of $\approx$2)
between about 13 and 23kpc. We nominally identify the MAMBO source with
29a but note that all three objects are part of a complex structure
with strongly varying colors reflecting different obscurations, a unique
object in our sample.

\paragraph{MMJ120530-0747.7 (07, u) } A faint blue object coincides
with a faint, uncertain radio source close to the edge of the search
radius.\footnote{At the chosen scaling when making the color images,
the red channel appears noisier than the blue and green channels. Thus,
we see enhanced ``red'' noise in the color image.} It is the only
candidate counterpart which is not detected in the K-band and the
bluest in our sample. The photometric redshift of 0.35 grossly differs
from the radio/mm estimate ($\sim$2.8), opening the possibility of the
identification being incorrect and the object a foreground dwarf. The low
mass of the relatively faint foreground object ($M_{K}\gtrsim -19$) makes
a lensing effect less likely. The probability of $\sim$0.1 for a chance
coincidence \citep[in this case based on the B magnitude and the B counts
of][]{pozzetti98} is consistent with finding such a case in our sample.

\paragraph{MMJ120531-0748.1 (04, u)}
The identification in the optical/near-infrared is not obvious. The faint
and uncertain radio source is 4.4\arcsec\ from the nominal MAMBO
position, behind and 1.0\arcsec\/ from the center of an extended
optical/near-infrared bright source (04a) with a photometric
redshift estimate of 0.25.  This is far below the radio/mm
estimate. Galaxy-galaxy-lensing could hence play a significant role
in MMJ120531-0748.1 --- the radio position and the closest position on
the estimated Einstein ring radius differ by less than 1$\sigma$ of the
radio positional error. The putative lensed object must be K$\gtrsim$18.
The morphology could have similarity to SMMJ04431+0210 
\citep{smail99,frayer03}, buth with a smaller separation making strong 
lensing more likely.

\paragraph{MMJ120534-0738.3 (42, g)}
This is the other of the two cases where two radio sources are detected
within our search radius. The identification of the radio counterpart
seems relatively clear, however, since a robust brighter radio source is
only 2.1\arcsec\ from the MAMBO position while the second is more distant
(5.5\arcsec\/) and is both fainter and of uncertain reliability.  We adopt
the first one, which is located inside a group of three faint objects
(42a-c) that are 1.3 - 2.0\arcsec\ (1.6 to 3$\sigma$) from the radio
position. An identification with 42a (z$_{phot}=$1.52$^{+0.24}_{-0.34}$)
or 42b (z$_{phot}=$2.83$^{+0.67}_{-0.24}$) is possible, the radio/mm
redshift estimate (z$_{CY}\sim$2.1) is consistent with the photometric
redshifts of either source. We adopt in the following an identification
with the closest object 42a but consider an identification with b (or a
blank field) possible.  Source No. 42 appears extended in the MAMBO map,
the peak flux (Table~\ref{tab:assoc}) thus underestimates the total flux.
The source is in a border region between two K-band fields which may
affect the quality of the K image and the photometric redshifts.

\paragraph{MMJ120539-0745.4 (16, g)} 
While the radio counterpart is faint, its position is confirmed by the
PdBI mm detection of \citet{dannerb02} which we adopt as best position.
It is still a blank field in deeper images (K$>$22.7mag, Lehnert et
al. 2004, in preparation).\footnote{The noise peaks in the K-band seen
in the color image are due to overlaping regions between two separate
pointings in our K-band mosaic.}

\paragraph{MMJ120545-0738.8 (40, u)}
A 40$\mu$Jy  uncertain radio source is located 3.9\arcsec\/ west of the
nominal bolometer position. The faint object 40a is within 1.6$\sigma$
of the radio positional error from the radio position.  Directly at the
MAMBO position and 4$\sigma$ from the radio peak, however, a compact
elliptical shaped extremely red object (40b) with a photometric redshift
z$=$1.52$^{+1.24}_{-0.25}$ and a best-fitting burst-like SED is seen.
An identification with this object may also be possible, its radio/mm
index could still be consistent with the suggested photometric redshift
range given the noise level of the VLA map. The observed radio peak
would then be unrelated. Additional MAMBO on-off observations in winter
2001/2002 could not conclusively settle this ambiguity but showed
the strongest signal at the position of the ERO. Only deeper radio
interferometry or mm-interferometry can clarify this ambiguity, in the
following we adopt an identification with the ERO which is another high
J-K object.

\paragraph{MMJ120546-0741.5 (31, g)}
This source was discussed by \citet[][]{dannerb02} and is a blank
field at the depth of the images presented here. Lehnert et al. (2004,
in preparation) detect a K-band counterpart (K$_{s}$=21.9mag) in VLT
imaging. \citet{eales03} detect the source at 850$\mu$m and suggest a
submm/mm redshift estimate consistent with the radio/mm estimate or with
significantly higher redshifts.

\subsection{Photometric redshifts vs. radio/mm redshift 
estimates}

Figure~\ref{fig:photvscy} compares the photometric redshifts with the
radio/mm redshift estimates. We consider this mostly a plausibility
check given the uncertainties in both methods.  The sources are often 
faint with relatively large photometric uncertainties in some
of the band-passes (especially the optical ones), and the templates
used in the photometric redshift estimates may not be fully appropriate 
for the likely ULIRG-like SEDs. The radio/mm redshift estimates are
affected by the low significance of the radio and mm detections
and the internal scatter of the relation for the radio/mm redshift estimate.
It is reassuring that most sources are consistent with the main diagonal 
in this diagram, with few well defined exceptions. Two sources (07 and 04)
are at low photometric redshift but high radio/mm estimate. We have argued
above for the possibility of foreground objects in these two cases. One
object (36) has a relatively low radio/mm estimate which could indicate
AGN contribution but is still consistent with scatter in this relation.
The optical `blank fields' tend to have high radio/mm redshift estimates.
This is clearly consistent with them being faint/obscured high redshift 
sources, but mis-identifications due to the relatively uncertain radio 
sources may also contribute. 

\section{Optical/near-infrared properties: very faint counterparts}
\label{sect:optnir}

The optical/NIR counterparts detected for the NDF mm sources are on
average faint even compared to published values for submm galaxies. This
is illustrated in Fig.~\ref{fig:magdist} which shows the distributions
of I and K band magnitudes and limits for our sample as well as for
three samples of submm sources where these magnitudes are available for
identifications obtained with the help of radio or mm interferometry. For
consistency with our sample, we did not consider from those papers
the more unreliable identifications solely based on the bolometer
survey positions. The first of these surveys is the SCUBA 8mJy survey
\citep{scott02,ivison02} which, for plausible assumptions on redshift and
SED of the objects, should be roughly similar in depth to our data. The
median submm flux of the radio-identified 8mJy sources is S$_{850\mu
m}\sim$8.3mJy, while the median mm flux of our radio-identified
sources is S$_{1.2mm}\sim$3.0mJy. This ratio $\sim$2.8 of 850$\mu$m
and 1.2mm fluxes is consistent with z$\sim$3 ULIRG-like SEDs and is
borne out in studies of well-detected robust submm and mm sources
\citep[e.g.,][]{hughes98, ivison98, gear00, ivison00, lutz01}.  We have 
augmented data for
the second set, the SCUBA cluster lens survey \citep[][and references
therein]{smail02}, with two more objects behind one of their clusters
detected by \citet{cowie02} and discussed by \citet{ledlow02}. For one
of these we adopt the revised identification of \cite{neri03}. Finally,
we use magnitudes from the CUDSS survey, for radio identified objects
in the 3h and  14h fields \citep{webb03a,webb03b} including the object
studied by \citet{gear00}. The latter two surveys differ more from our
survey in strategy (cluster lens) or depth (CUDSS) but may still serve
as comparisons.

As shown in Fig.~\ref{fig:magdist}, the NDF mm sources are significantly
fainter both in K and in I. Taking the median of the observed magnitudes
or limits, we find K$\sim$21.5 and I$\sim$25.5 for the mm objects
versus K$\sim$19.6 and I$\sim$23.3 for the identifications from the 8mJy
survey. The cluster lens and CUDSS objects have smaller statistics but
show intermediate counterpart brightnesses \citep[see also the recent deep
SCLS imaging of][sometimes finding very faint counterparts]{frayer04}.
Eliminating all those NDF identifications based on radio sources that we
considered more uncertain above does not change the values strongly - the
median of the remaining objects is K$\sim$20.9 and I$\sim$24.9, i.e. the
counterparts of the MAMBO sources are about 2 magnitudes fainter than
those in the 8mJy 850$\mu$m survey. One major difference between the two
populations is the absence in our sample of bright K$\sim$18 counterparts,
of which several are found for the 8mJy survey.\footnote{We note that
different aperture sizes have been used in making this comparison.
For example, \citet{ivison02} use a 3 arc second diameter aperture to
estimate their fluxes intead of our 2 arc second apertures.  In sources
with complex extended morphologies or nearby objects, the larger apertures
could lead to systematically brighter magnitudes.  However, it seems
unlikely that this alone could lead to an offset of 2 magnitudes.} The
trivial explanation of such a difference, our sample just representing
much fainter members of the same population, is clearly inconsistent
with the depths and areas of the various surveys. We also do not see
a trend in our sample for bright mm sources being brighter in K, as
possibly present in the SCUBA cluster lens survey \citep{smail02}.

\subsection{Colors and the relation to field galaxies and the extremely red 
population}

Another way to represent the optical/near-infrared properties
is in the form of an I-K vs. K magnitude-color diagram as
presented for SCUBA sources in \citet{smail02}, \citet{ivison02},
and \citet{webb03a,webb03b}. Again, we restrict our analysis
(Fig.~\ref{fig:ik}) to objects with interferometric positions. For
comparison, the figure also includes the field galaxy population from
our NDF data and the expected properties of objects with SEDs similar
to local ULIRGs. For these, we have picked both an object that is red
in the UV/optical (Arp 220) and a blue one (IRAS 22491-1808) from the
objects studied in the UV by \citet{trentham99} and \citet{goldader02} and
completed its SED with far-infrared data from \citet{klaas01} and large
aperture optical/near-infrared data from elsewhere in the literature. The
color-magnitude relation is then derived for these redshifted SED {\em
shapes} but scaling the absolute flux to an observed flux of 5mJy at
1.2mm. The main results from this analysis are the following: First,
the optical/near-infrared colors of (sub)mm galaxies scatter cover a wide
range at a given magnitude \citep[the `diversity' of SCUBA galaxies,][]
{ivison00}. Second, with few exceptions this range is within the
envelope expected for the range of redshifted local ULIRG SEDs. These
exceptions are from the 8mJy sample which generally has a surprisingly
large scatter of I-K colors, larger than for the mm counterparts, and
includes some very blue objects. In our sample we find only one object
with a limit consistent with I-K$<$3.3 and one measurement I-K=3.33,
while there are about half a dozen identifications in the 8mJy sample
with objects at I-K$<$3.3, a color well in the range of the normal
field population.  Third, for K$>$19, the average (sub)mm source is
redder in I-K than an average field galaxy of the same magnitude. This
gives some support to the suggestion that nearby EROs are the most
likely counterparts to SCUBA or MAMBO sources, but the dispersion in
colors of both field galaxies and (sub)mm sources is too large to make
this a robust criterion for identifications of individual submm sources
without accurate positions. More specifically, the brightest (sub)mm
counterparts (K$<$19) are statistically indistinguishable in I-K colors
from the field population while (sub)mm counterparts are significantly
redder at 19$<$K$<$21. It is difficult to quantify at this point how the
trend continues at  K$>$21 since observational limits in both K and I are
significant for several of the samples involved. While the dispersion
of I-K colors is larger for the 8mJy survey counterparts than for our
objects we do not see a significant difference in the mean color at a
given magnitude.

We compare the J-K colors which are available for our full sample with
expectations for redshifted SEDs of ULIRGs. Fig.~\ref{fig:jk} shows the
expected J-K colors as a function of redshift for a sample of six local
ULIRGs, using their rest-frame UV to NIR SEDs assembled on the basis of
the UV data of \citet{trentham99} and \citet{goldader02} and of large
aperture optical/NIR photometry from the literature. For this type of
SEDs, the J-K color is not a unique indicator of redshift, but excursions
to large J-K colors can occur for redshifts above 2 and red SEDs like
that of Arp220. Like Fig.~\ref{fig:ik}, this diagram shows the large
variety of colors for local ULIRGs. The locations of the MAMBO sources
spread from J-K$\sim$2, corresponding to intrinsically bluer SEDs or
lower redshifts, to J-K$>$3.4 which is expected only for red Arp220-like
SEDs and high redshifts. Again, the most likely interpretation is that
the spread in the intrinsic optical/UV SEDs of mm galaxies is large,
similar to the spread in local ULIRGs.

At least 5 of our 18 sources have ERO (I-K$>$4) counterparts. Given
the long standing discussion about the contributions of passively
evolving ellipticals and of dusty starbursts to this class, it is of
interest to place our objects in this context. Fig.~\ref{fig:pozzetti}
shows an I-K vs. J-K color-color diagram proposed by \citet{pozzetti00}
to address this issue.  Based on the dissimilarity between a 4000\AA\
break and a reddened more smooth starburst continuum, this diagram is
expected to separate the two categories for redshifts in the range of
roughly 1 to 2. With the possible exception of MMJ12507-0748.1 (No. 03)
the proposed mm and submm counterparts avoid the region of z$\sim$1-2
evolved populations, which might be found in case of false identifications
of mm sources with such objects. It is not possible to draw the inverse
conclusion from this diagram, i.e. to conclude that we are observing
starbursts at z$\sim$1-2. The region populated by the (sub)mm sources
is indicative of smooth spectra which are not very specific on redshift,
or of spectra with a break between J and K. This caveat is particularly
relevant since some (sub)mm objects are significantly fainter than the
z$\sim$1-2 EROs for which this method was developed.

Recently, \citet{franx03} presented a technique to select optical-break
galaxies at a redshift z$>$2 with the near-infrared filters J$_{s}$
and K$_{s}$. \citet{vandokkum03} reported spectroscopic redshifts of a
handful of such (J$_s$-K$_s$)$>$2.3 pre-selected galaxies, five of six
indeed lying at redshifts higher than z$>$2 (z=2.4--3.5). Individual
similar objects have been found in other surveys \citep[e.g.,
][sometimes called `HEROs']{dickinson00,maihara01, totani01a,im02}. We
have probed for an overlap between this population and the MAMBO
identifications. \citet{franx03} report a surface density of $3\pm
0.8$ arcmin$^{-2}$ for objects with J$_s$-K$_s$$>$2.3 and K$_s<$22.5.
In our sample from a survey area of roughly 100 square arcmin we suggest 3
identifications with J-K$>$2.3, K$<$22 galaxies. Given several more J$>$24
limits, a few more might be found in deeper images to K=22.5 among the
objects with current K$>$22 limits, and among the MAMBO sources without
radio identification that are not discussed in this paper. It is thus
fair to assume that the contribution of mm galaxies in the few mJy range
to the population of luminous red galaxies as observed by \citet{franx03}
is of the order one to a few percent. This small overlap is consistent
with submm/mm followup for the four HEROs of \citet{totani01a}, reporting
very weak sub/mm signals \citep[][and Andreani et al. in prep.]{coppin02}.
Also, only one of the six J$_s$-K$_s$$>$2.3 objects with spectroscopic
redshifts reported by \citet{vandokkum03} has a spectrum consistent
with that of a dusty starburst.  On the other hand, the fraction of red
J-K sources (J$_s$-K$_s$$>$2.3 and K$_s<$22.5) in the mm population is
significantly  higher.  Of our MAMBO sources with radio counterparts,
at least 3 out of 18 ($\sim$20\%) are identified with red J-K sources.
It may be higher, but the lack of deeper J-band data prevents us
from estimating the colors of the faintest of the K-band counterparts.
There might be an evolutionary connection beyond this small overlap, i.e.
a contribution of mm galaxies in quiescent or post-active phases to the
population of luminous red galaxies/HEROs.

\subsection{Dependence on Radio Properties: Radio-Pre-Selection
and Detection Limits}

One of the major issues in our developing understanding of the nature
of (sub)mm emitting galaxies at high redshift is how does the method
of selection influence the properties of the sources?  For example,
targeted submm follow-up of optically faint radio sources \citep[OFRS:
$S_{1.4GHz}\geq 40\mu$Jy, I$>$25;][]{chapman01} has been used as an
efficient method to detect and study a significant `radio pre-selected'
fraction of the (sub)mm population.  This is a particularly important
comparison for our study.  Because we have limited ourselves to a
investigation of only those mm sources in the NDF with interferometric
identifications, we might expect a radio-preselected sample to have
similar properties.  Indeed, the strength of the conclusions from such
studies obviously depends on the amount of overlap with the population
detected in unbiased `blind' surveys (i.e., those not dependent of
radio pre-selection), and on the nature of the biases induced by
the OFRS selection. Both \citet{barger00} and \citet{chapman01}, in
comparing the submm detection efficiency of radio sources with that
of blank field surveys, report a recovery rate of about 70\% of bright
($>$5-6mJy) submm sources when targeting $S_{1.4GHz}\geq 40\mu$Jy radio
sources with very faint near-infrared or optical counterparts (HK$^\prime
>$20.5 or I$>$25, respectively).  In contrast, only 10\% of the refined
8mJy sample are reported to have an $S_{1.4GHz}\geq 40\mu$Jy, I$>$25
counterpart \citep{ivison02}.  Using the same criteria, we recover 10
of 42 (about 25\%) of the original NDF MAMBO list.

While the recovery fraction of optically faint radio identified
sources presented here is similar to the low 8mJy recovery rate of
\citet{ivison02} at first glance, a clear difference exists in the
nature of the sources not matching the $S_{1.4GHz}\geq 40\mu$Jy, I$>$25
criteria. We find 7 to 8 (about 20\%) optically bright (I$<$25) radio
sources, while \citet{ivison02} find 12 of 30 optically bright sources
(40\%) in their refined sample with $S_{1.4GHz}\geq 40\mu$Jy radio
counterparts.  This is another manifestation of the clear difference in
typical counterpart brightnesses between the NDF MAMBO sources and the
8mJy survey reported above. Irrespective of these differences, a large
recovery rate by the radio pre-selection technique for faint optical
sources is not supported by both `blind' surveys.

Part of the difference between the suggested 70-75\% recovery rates
and the lower numbers found in the blind surveys may be simply due
to reference integral counts of the complete submm population and due
to small areas that we used in these surveys.  The counts adopted by
\citet{barger00} and \citet{chapman01} are more towards the low end of
the spread of the count determinations summarized in \citet{blain02}
and the number of sources in these field meeting their criteria was
relatively small (for example, in \citet{chapman01}, the recovery rate
was determined based on 11 radio sources with submm detections and
2 sources without radio counterparts). In any case, a large recovery
rate of the optically faint radio pre-selection is not ensured, and the
selection effects are complex \citep[see also][]{chapman02a}.  While the
radio detection will overall favor low redshift \citep{carilli99},
optical faintness will prefer among those objects the higher redshifts
ones. This is indeed suggested in our radio-identified sample where the
mean redshift (approximated by the radio/mm indicator) is $<z>=2.8\pm 0.2$
for the I$>$25 objects compared to $<z>=2.4\pm 0.2$ overall.

While a comparison of the recovery rate of radio pre-selected sources
with faint optical counterparts is interesting and instructive,
perhaps a more robust comparison can be made with \citet{chapman03c}.
In \citet{chapman03c}, the radio pre-selected sample is defined
without imposing constraints on magnitudes of the optical/near-infrared
counterparts to the radio sources, now finding noticeable numbers of
optically brighter (I$<$25) radio-identified submm sources in contrast to
the earlier radio-preselected studies.  Overall, they recover about 70\%
of all submm sources which is formally higher than both the MAMBO source
recovery rate for our survey and that of \citet[][which have recovery
rates of $\sim$40\% and $\sim$60\% respectively]{ivison02}.  Similar to
the differences we have found between our results and those of the 8 mJy
SCUBA survey of \citet{ivison02}, we also find a substantial difference
between our results and the sources meeting our selection criteria of
S$_{850\mu m}\sim$8mJy and $S_{1.4GHz}\geq 40\mu$Jy in \citet{chapman03c}.
For the 10 sources in \citet{chapman03c} meeting these criteria, we find
a median I-band magnitude of $\approx$24 with only 20\% of the sources
having I-band magnitudes fainter that I=25 (2 out of 10).  Such a result
is in stark contrast to our results with a median I-band magnitude of
$\sim$25.5 and $\sim$80\% of sources having I$>$25.

It is important to note that the differences in recovery rates and
counterpart properties are not a direct consequence of the applied
radio flux limits combined with strong trends with flux. Making a
higher flux cut (40$\mu$Jy as in our sample) in the radio  detections
of \citet{ivison02} only removes a small number of sources from their
survey (3 or 4 depending on how to treat a binary radio source, leaving
a recovery rate of still $\sim$50\%).  Similarly, the majority of sources
in the radio preselected samples of \citet{barger00}, \citet{chapman01},
and \citet{chapman03c} have $S_{1.4GHz}\geq 80\mu$Jy (well above their
detection limit), changing the lower radio flux to our value thus has
little effect .

\section{Optical to radio SED constraints}
\label{sect:optradsed}

We discuss the near-infrared (rest frame optical) to radio spectral
energy distributions of our sources using an updated version of the
diagram presented in Paper~I, combining a spectral index between the
K band and the submm and the well known radio/submm spectral index
(Fig.~\ref{fig:rgdiag}). To be able to compare MAMBO and SCUBA sources,
we estimate 850$\mu$m fluxes for those MAMBO sources where they have not
been measured, by scaling the 1.2mm fluxes by a factor of 2.5 which is
consistent with the median value of published ratios. This implicitly
assumes that there is no major difference between the submm and mm
populations. In Fig.~\ref{fig:rgdiag}, the MAMBO objects are also on
average below the SCUBA ones, consistent with their optical/near-infrared
faintness derived in the previous section.

Given the opposite trends in K corrections for the (sub)mm flux on one
side and the radio and optical/NIR fluxes on the other side, redshifting
a dusty galaxy SED  will move the object from top left to lower right in
this diagram.  Such a trend is indeed indicated in the upper panel. The
lower panel verifies and helps to visualize this expectation by plotting
the loci for the redshifted SEDs of six local ULIRGs for which UV to radio
SEDs are available. These clearly provide an indication of the scatter
in the properties of the local population. In particular, significant
offsets to the left are possible for objects with strong AGN, such as
IRAS 19254-7245 with its reddened Type 2 AGN \citep[][]{mirabel91} and
strong radio emission.  It will be interesting to probe for AGN activity
for the objects in the left ($\alpha$(850$\mu$m,1.4GHz)$\lesssim$0.65)
part of the diagram, indeed some of the objects there show evidence for
AGN activity (LE850.12, \citet{ivison02}, J02399-0134, \cite{soucail99}).
For the radio/submm spectral index on the horizontal axis the dispersion
for a larger sample is of the order 0.16 \citep{carilli00a}, for the
NIR/submm index of similar magnitude given the scatter arising from
adopting different ULIRG SEDs (Fig.~3 of Paper~I.). A similar scatter
is observed in the high redshift (sub)mm population, shown by labeling
in the top panel objects with CO-confirmed redshifts with the redshift
value. Four of these six objects may also serve as an indication of
a locus of z$\sim$2.5 objects that is independent of the uncertain
applicability of local templates.  The average locus of these z=2.4
to 2.8 (mean z=2.6) objects is $\alpha$(850$\mu$m,1.4GHz)=0.76 and
$\alpha$(2.2$\mu$m,850$\mu$m)=-0.99.  Better statistics from on-going CO
follow-up observations providing confirmed redshifts is highly desirable
to improve this calibration of Fig.~\ref{fig:rgdiag}.  Although,
the CO confirmed redshift may introduce an upward bias for this point
if the successful optical redshift measurements on which they depend
on correspond to brighter counterparts on average.  Less bias should
be present in the horizontal (radio/submm) direction. We find about
two thirds of the MAMBO/SCUBA objects towards the `high redshift'
(z$\gtrsim$2.5) direction to the right of this point (43/69 at
$\alpha$(850$\mu$m,1.4GHz)$>$0.76). This is an underestimate, since the
plot shows almost exclusively radio detected objects, thus missing roughly
one-third to a half of the parent samples including the potentially
highest redshift objects. Altogether, this indicates that the median
redshift of 2.4 suggested for the bright SCUBA population from optical
redshifts \citep{chapman03a} is an underestimate of the median redshift
of the complete population.

A conclusion drawn in Paper~I from this type of diagram is that MAMBO
and SCUBA sources are at high redshift and/or are at least as obscured
as local ULIRGs.  A technical factor affecting this reasoning is due to
the unknown spatial structure of the objects. If the objects were very
large and considering the effects of cosmological surface brightness
dimming, the observed small structures (Fig.~\ref{fig:idplot1}) may just
reflect the central high surface brightness regions of more luminous
extended objects that cannot be fully retrieved at realistic noise
levels by just increasing aperture size. This would cause a tendency
to underestimate the total near-infrared flux. There is evidence for
some objects of that type in the (sub)mm population from near-IR imaging
\citep[e.g., Lockman 850.1,][]{lutz01}, and radio mapping \citep{ivison02}
finds some objects resolved by a 1.4\arcsec\ beam. Examples of extended
and complex morphologies are also seen in HST images of radio-preselected
submm galaxies \citep{chapman03b}. For many other well studied (sub)mm
sources the detected regions fit into 2\arcsec\ apertures but this may
be deceiving effects of surface brightness dimming if they are at very
high redshift.  Some estimate of the magnitude of the effect can be
gained from comparison to the local ULIRG population.  Half light radii
have been estimated at several wavelengths, with median results of the
order 5.3 kpc in U \citep{surace00}, 3 kpc in I \citep{zheng99} and about
1.5 kpc in H and K \citep{colina01,tacconi02}.  At redshifts around 3 an
aperture of 2\arcsec\ diameter spans about 15 kpc. Half light radii of
$\sim$4 kpc in the rest frame V band (observed K band) would then mean
no significant loss of light outside our aperture. This is only true as
long as the structural analogy to local ULIRGs holds, of course, and a
major fraction of light could be lost for objects more than a factor of 2
larger, especially if at z$\gtrsim$4. Such large objects would be placed
too low in Fig.~\ref{fig:rgdiag} by our measurements.  This implies
that the displacement of many SCUBA and MAMBO objects down-wards from
the location of local ULIRGs could include an effect of large spatial
extent as well as obscuration of the rest frame optical part of the
SED.  Testing such a scenario would require extremely deep K imaging.
Size constraints at other wavelengths (optical, mm, radio) will be useful
as well, but for this question they will have to be interpreted with
some caution considering the strong trends for local infrared galaxies,
where half light radii decrease as wavelength increases.

\section{What causes the difference between SCUBA and MAMBO populations?} 

The difference in brightness of about two magnitudes between typical 
MAMBO and SCUBA counterparts is a surprising result, given the selection of 
these objects at fairly similar wavelengths and on a part of the rest 
frame spectral energy distribution that does not show prominent breaks or 
features. We discuss in the following several potential contributors to 
this difference.

Given the faintness of the targets in comparison to instrumental
sensitivities at {\em all} wavelengths, the identification process
may still contain errors. For the NDF MAMBO sources, some of the
near 3$\sigma$ radio sources are clearly uncertain and may lead
to uncertain identifications. We have argued above, however, that
eliminating the uncertain sources as classified in \S~\ref{sect:assoc}
does not change the magnitude distribution of the MAMBO counterparts
significantly. For the SCUBA sources similar problems may be
present at the limit of the radio data. In addition, there are by
now examples where intensified identification efforts have rejected
previously adopted bright counterparts and lead to a significant upward
revision of the proposed counterpart magnitudes, the most notable being
HDF850.1 \citep{hughes98,richards99,downes99,dunlop02}, SMMJ00266+1708
\citep{frayer00} and SMMJ09431+4700 \citep{neri03}. Taking away a few of
the brightest or bluest 8mJy counterparts from Fig.~\ref{fig:magdist}
does not fully remove the difference to the MAMBO counterparts,
though.  Noting all these uncertainties we do not see any compelling
evidence for systematic mis-identifications.  Identification
uncertainties are also related to the presence of `complex' objects
with several components, e.g., one of the  mm galaxies discussed by
\citet{bertoldi00}. \citet{ivison02} emphasize the role of such objects
in the 8mJy survey. Again, the example of SMMJ09431+4700 \citep{neri03}
where the submm and CO source H7 is 4\arcsec\ from the object H6 with a
very similar (optical) redshift indicates that such grouping or clustering
exists. In such cases, the (sub)mm source could relate to just one or to
all of the components. We find only three such objects in our sample. For
No. 10 we have preferred one component because of photometric redshift
arguments. For objects 29 and 42 there is more uncertainty on whether
to identify the source with one or several components, but even the sum
of the components will not be a `bright' counterpart.

Optical/near-infrared magnitude differences could also reflect
a significant difference in the redshift distribution of the two
populations.  The effect in the optical/near-infrared will obviously
depend on the intrinsic SEDs, for ULIRG-like SED shapes and keeping
the (sub)mm flux fixed a difference in median redshifts of the order
$\Delta$z=1-2 may be necessary (Fig.~\ref{fig:ik}).  \citet{chapman03a}
suggest a median redshift of 2.4 for bright submm galaxies from optical
spectroscopy. Some of these optical redshifts still require confirmation
by CO detections, and it is natural to assume that the optical redshifts
are still biased towards the optically bright and low redshift end
of the population. This assumption is supported by SED arguments
(\S~\ref{sect:optradsed}).
Even at redshifts around or somewhat above 3, however, both SCUBA and
MAMBO observations effectively sample the long wavelength side of the
rest frame far-infared SED peak, unless the intrinsic dust temperatures
were extremely cold. Evidence for both wavelengths sampling the long
wavelength side is indeed present in many of the best studied SCUBA
sources \citep[e.g.,][]{ivison98,downes99,gear00,ivison00,lutz01}.
At such redshifts, it will be difficult to construct a redshift
distribution that fully accounts for the observed differences.  Redshift
estimates for the SCUBA population from the radio/submm spectral index
\citep{carilli00a}, and from estimates trying to more fully include
the rest-frame far-infrared part of the SED under the assumption of SEDs
similar to local ULIRGs \citep{yun02,wiklind03,aretxaga03} indicate median
redshifts in the range 2.5 to 3.5 with the possibility of a significant
high redshift tail \citep{wiklind03,aretxaga03}.  However, it is important
to note that such analysis suffer from possible template mismatches
between high and low redshift sources and using an unrealistically narrow
range of SED types and properties when making redshift estimates.
For example, \citet{aretxaga03} may perhaps put undue weight on single
templates by matching one template to each source to constrain the
possible redshift range.  This means that low redshift SEDs they have
in their analysis must be good analogues for high redshift sources.
In addition, a substantial number of higher redshift z$>$4 sources in the
MAMBO population is also suggested by the 850$\mu$m/1.2mm ratios presented
by \citet{eales03}, albeit with large uncertainties due to the technical
difficulties of the method. For the objects overlapping with our study
(Nos. 1, 3, 16, 25, 31) their 850$\mu$m/1.2mm redshift estimates can be
compared to the detection in various optical pass-bands - if they are at
high redshift, these objects should be undetected dropouts in optical
bands.  With the possible exception of No.3 (best submm/mm estimate
z$\sim$4.25 but a marginal detection in B) the results are consistent,
and even for this object considering the uncertainties. In summary,
while there may be a high redshift tail of the (sub)mm population
preferentially sampled by MAMBO, we find it difficult to explain the
magnitude differences by a bulk redshift difference.  This view is
consistent with Fig.~\ref{fig:rgdiag} which does not only show the
`diagonal' offset between the two sets expected in such a case.

Because of the degeneracy of redshift and dust temperature in
interpretations of the submm/radio spectral energy distributions
\citep[e.g.][]{blain03}, similar arguments can be made with respect
to a possible separation of the MAMBO and SCUBA populations by dust
temperature, for example due to an evolutionary difference between MAMBO
and SCUBA objects. One should note, however, the need for significant
obscuration in the rest frame optical when invoking `cool' rest-frame
FIR SEDs (e.g., Paper I).

The results shown here indicate that there are real differences in the
counterpart brightnesses and perhaps redshifts of the SCUBA and MAMBO
populations. On the sole basis of the SED arguments presented it is
difficult to robustly weight the contributions of real differences in
redshifts and spectral energy distributions of the mm and submm selected
populations to this difference, in comparison to caveats due to small
sample statistics, cosmic variance, and identification difficulties.
Significant progress in settling these issues will come from tests of
proposed identifications with available optical redshifts through CO
measurements \citep[e.g.,][]{neri03}, and from future direct submm/mm
redshift measurements for the fainter part of the population, most likely
through wide band searches for CO emission.

\section{Conclusions}

We have discussed optical/near-infrared identifications for those 18
of 42 sources in our MAMBO 1.2mm map of the NTT Deep Field region for
which interferometric positions are available through a VLA 1.4GHz
map and in three cases IRAM PdBI mm interferometry. In addition to
being the basis for identifications, our deep BVRIzJK imaging allows
the derivation of optical/near-infrared photometric redshifts for the
counterparts and for nearby objects. In comparison with radio/submm
redshift estimates, these photometric redshifts suggest that two of the
optical/near-infrared sources close to interferometric positions are in
the foreground, in one case likely lensing the background mm source. One
strongly lensed object in this sample is consistent with expectations
for the (sub)mm population \citep{blain99,chapman02c}. This leaves us
with eleven detections of counterparts at magnitudes of K=19 to 22.5, and
seven limits or blank fields with most limits at K$>$22. The I-K and J-K
colors of the counterparts are consistent with redshifted SEDs similar to
local ultra-luminous infrared galaxies, and likely with a similarly large
spread of the rest frame UV/optical SED properties. The counterparts of
mm sources contribute to the recently discussed population of J-K$>$2.3,
K$<$22.5 high redshift galaxies, but only at the few percent level for
the current mm survey depths.  At least $\sim$20\% of the MAMBO sources with
radio counterparts have J-K$>$2.3 and K$<$22.5.

The counterparts to NDF mm sources are on average about 2 magnitudes
fainter than counterparts presented for the 8mJy 850$\mu$m survey
which is of similar depth and are similarly faint compared to radio
pre-selected submm sources. Remaining mis-identifications, redshift or
temperature differences between the two populations, and small number
statistics/cosmic variance all may contribute to this difference,
at levels that are hard to quantify from currently available data. Our
result reinforces the view that direct (e.g., wide band CO) spectroscopic
redshifts may be necessary for a substantial fraction of the (sub)mm
population which is very faint in the optical/near-infrared. From a
comparison of near-infrared/submm/radio spectral indices with those
of submm sources with CO-confirmed redshifts we suggest that the
fraction of (sub)mm galaxies at z$>$2.5 is about two thirds for
the interferometrically located ones and larger when adding the
radio-undetected part of the population.

\acknowledgements{We would like to thank the staffs of ESO, IRAM, and VLA
for their support with the data acquisition and reduction. We thank Andrew
Baker for helpful discussions.  We also acknowledge the contributions of
the referee in clarifying a number of important points.  This research has
made use of the NASA/IPAC Extragalactic Database (NED) which is operated
by the Jet Propulsion Laboratory, California Institute of  Technology,
under contract with the National Aeronautics and Space Administration. }

\clearpage
\begin{deluxetable}{rrrrlrrrrrcl}
\tabletypesize{\tiny}
\tablecolumns{12}
\tablewidth{0pt}
\tablecaption{NDF MAMBO sources and associated 1.4GHz radio sources}
\tablehead{
\multicolumn{5}{c}{MAMBO 1.2mm properties}&\multicolumn{6}{c}{VLA 1.4GHz
properties}&\\
\colhead{Source}&\colhead{No}&\colhead{RA}&\colhead{DEC}&
\colhead{S$_{1.2mm}$}&\colhead{RA}&\colhead{DEC}&\colhead{S$_{1.4}$}&
\colhead{Sep}&\colhead{P}&\colhead{Q}&\colhead{Comment}
\\
\colhead{}&\colhead{}&\colhead{(J2000)}&\colhead{(J2000)}&
\colhead{mJy}&\colhead{(J2000)}&\colhead{(J2000)}&\colhead{$\mu$Jy}&
\colhead{arcsec}&\colhead{}&\colhead{}&\colhead{}
\\
\colhead{(1)}&\colhead{(2)}&\colhead{(3)}&\colhead{(4)}&
\colhead{(5)}&\colhead{(6)}&\colhead{(7)}&\colhead{(8)}&
\colhead{(9)}&\colhead{(10)}&\colhead{(11)}&\colhead{(12)}
}
\startdata
MMJ120507-0748.1 & 03 & 12:05:07.96 & -07:48:11.9 &  4.6$\pm$1.0 & 12:05:08.12$\pm$0.03 & -07:48:11.6$\pm$0.6 &  88 & 2.4 & 0.009 & g & \\
MMJ120508-0743.1 & 26 & 12:05:08.36 & -07:43:06.8 &  2.7$\pm$0.7 & 12:05:08.66$\pm$0.06 & -07:43:05.6$\pm$1.2 &  42 & 4.6 & 0.107 & u & \\
MMJ120509-0740.0 & 34 & 12:05:09.75 & -07:40:02.5 &  3.3$\pm$0.6 & 12:05:09.80$\pm$0.05 & -07:40:06.6$\pm$1.1 &  47 & 4.1 & 0.072 & u & \\
MMJ120510-0747.0 & 10 & 12:05:10.97 & -07:47:00.4 &  3.1$\pm$0.8 & 12:05:10.88$\pm$0.06 & -07:46:57.6$\pm$1.3 &  41 & 3.1 & 0.074 & u & \\
MMJ120516-0739.4 & 36 & 12:05:16.21 & -07:39:27.1 &  2.2$\pm$0.6 & 12:05:16.06$\pm$0.03 & -07:39:25.6$\pm$0.5 & 459 & 2.7 & 0.002 & g & \\
MMJ120517-0743.1 & 25 & 12:05:17.93 & -07:43:06.9 &  4.3$\pm$0.6 & 12:05:17.88$\pm$0.06 & -07:43:09.6$\pm$1.3 &  40 & 2.8 & 0.070 & u & \\
 & & & & & 12:05:17.86$\pm$0.02 &-07:43:08.5$\pm$0.2& & & & &PdBI pos.\\
MMJ120519-0749.5 & 01 & 12:05:19.98 & -07:49:33.8 &  5.2$\pm$1.0 & 12:05:19.90$\pm$0.04 & -07:49:35.6$\pm$0.7 &  72 & 2.2 & 0.009 & g & \\
MMJ120520-0738.9 & 39 & 12:05:20.45 & -07:38:56.7 &  2.7$\pm$0.8 & 12:05:20.63$\pm$0.03 & -07:38:55.6$\pm$0.5 & 336 & 2.9 & 0.003 & g & \\
MMJ120522-0745.1 & 18 & 12:05:22.86 & -07:45:10.0 &  2.0$\pm$0.5 & 12:05:23.12$\pm$0.03 & -07:45:14.6$\pm$0.6 &  90 & 6.0 & 0.035 & g & \\
MMJ120524-0747.3 & 08 & 12:05:24.86 & -07:47:20.9 &  2.0$\pm$0.6 & 12:05:24.81$\pm$0.05 & -07:47:23.6$\pm$1.1 &  47 & 2.8 & 0.042 & u & \\
 & & & & & 12:05:24.54$\pm$0.05 & -07:47:20.6$\pm$1.1 &  48 & 4.8 & 0.082 & u & \\
MMJ120526-0746.6 & 13 & 12:05:26.85 & -07:46:41.8 &  2.6$\pm$0.6 & 12:05:26.76$\pm$0.06 & -07:46:40.6$\pm$1.3 &  41 & 1.8 & 0.035 & g & \\
MMJ120530-0741.6 & 29 & 12:05:30.17 & -07:41:41.2 &  2.3$\pm$0.6 & 12:05:30.25$\pm$0.03 & -07:41:45.6$\pm$0.6 &  85 & 4.5 & 0.024 & g & \\
MMJ120530-0747.7 & 07 & 12:05:30.85 & -07:47:43.9 &  3.0$\pm$0.8 & 12:05:31.07$\pm$0.06 & -07:47:39.6$\pm$1.3 &  40 & 5.4 & 0.134 & u & \\
MMJ120531-0748.1 & 04 & 12:05:31.41 & -07:48:07.2 &  2.7$\pm$0.9 & 12:05:31.13$\pm$0.05 & -07:48:05.5$\pm$1.0 &  53 & 4.4 & 0.058 & u & \\
MMJ120534-0738.3 & 42 & 12:05:34.81 & -07:38:20.2 & $>$2.2$\pm$0.9$^{\ast}$ & 12:05:34.89$\pm$0.04 & -07:38:18.5$\pm$0.9 &  61 & 2.1 & 0.012 & g & \\
 & & & & & 12:05:34.49$\pm$0.06 & -07:38:17.6$\pm$1.3 &  41 & 5.5 & 0.130 & u & \\
MMJ120539-0745.4 & 16 & 12:05:39.37 & -07:45:24.7 &  3.4$\pm$0.7 & 12:05:39.48$\pm$0.05 & -07:45:26.5$\pm$1.0 &  55 & 2.4 & 0.020 & u & \\
 & & & & & 12:05:39.47$\pm$0.02&-07:45:27.0$\pm$0.3& & & & &PdBI pos.\\
MMJ120545-0738.8 & 40 & 12:05:45.98 & -07:38:51.8 &  3.7$\pm$0.8 & 12:05:45.72$\pm$0.06 & -07:38:52.5$\pm$1.3 &  40 & 3.9 & 0.102 & u & \\
MMJ120546-0741.5 & 31 & 12:05:46.56 & -07:41:33.2 &  6.5$\pm$0.9 & 12:05:46.53$\pm$0.06 & -07:41:32.5$\pm$1.2 &  42 & 0.8 & 0.009 & g & \\
 & & & & & 12:05:46.59$\pm$0.02&-07:41:34.3$\pm$0.4& & & & &PdBI pos.\\
\tablecomments{
Col. (1) --- MAMBO source.\\ 
Col. (2) --- Short number on the original MAMBO NDF source list.\\
Col. (3)-(4) --- J2000 coordinates of the source in the MAMBO data.\\
Col. (5) --- MAMBO 1.2mm peak flux density. $^{\ast}$: The
source MMJ120534-0738.3 appears to be extended and may be multiple or 
smeared. For consistency with the other sources, we
list its peak flux which is effectively a lower limit to the true flux.  
The source was included because of a reliable detection in total flux
and a 1.4GHz counterpart.\\
Col. (6)-(7) --- J2000 coordinates of the associated radio source.\\
Col. (8) --- VLA 1.4GHz peak flux density. The average rms of the radio map is 13
 $\mu$Jy.
See also col. (11) for possible effects of striping residuals caused by 
bright sources contained in the radio map (see \S~\ref{sect:obsandred}). \\
Col. (9) --- Separation between MAMBO position and VLA position.\\
Col. (10) --- Probability that the association is a chance coincidence.\\
Col. (11) --- Quality flag for good (g) and uncertain (u) radio sources. To
 declare the quality of a radio source as good two criteria has to fulfilled:
 (1) source does not disappear in the 'background-subtracted' radio map, and
 (2) no eye-catching clustering of $>$40$\mu$Jy peaks in the adjacent region.
}
\enddata
\label{tab:assoc}
\end{deluxetable}
\clearpage

\begin{deluxetable}{lrrrllllllllr}
\tabletypesize{\tiny}
\tablecolumns{13}
\tablecaption{Optical/near-infrared objects close to the interferometric 
positions}
\tablehead{
\colhead{Source}&\colhead{RA}&\colhead{DEC}&\colhead{Sep}&
\colhead{B}&\colhead{V}&\colhead{R}&\colhead{I}&\colhead{z}&
\colhead{J}&\colhead{K}&\colhead{z$_{phot}$}&\colhead{P}\\
\colhead{}&\colhead{(J2000)}&\colhead{(J2000)}&\colhead{arcsec}&
\colhead{mag}&\colhead{mag}&\colhead{mag}&\colhead{mag}&\colhead{mag}&
\colhead{mag}&\colhead{mag}&             &             \\
\colhead{(1)}&\colhead{(2)}&\colhead{(3)}&\colhead{(4)}&
\colhead{(5)}&\colhead{(6)}&\colhead{(7)}&\colhead{(8)}&\colhead{(9)}&
\colhead{(10)}&\colhead{(11)}&\colhead{(12)}&\colhead{(13)}\\
}
\startdata
{\bf 03a} & 12:05:08.07 & -07:48:11.7 &  0.8 & $\approx$27.1 & $>$26.5 & !25.9 & $>$25.5 &   $\approx$26.1 &   !23.3 &   21.48 &   
        2.11$^{+0.26}_{-0.61}$ & 0.030\\
03b & 12:05:07.97 & -07:48:08.7 &  3.7 & $>$27.4 & $>$26.5 &   !25.8 &   $\approx$25.3 &   25.48 & $>$24.0 &   $\approx$21.9 &   
        3.64$^{+0.84}_{-0.50}$ \\
03c & 12:05:08.03 & -07:48:17.5 &  6.1 &   24.93 &   24.49 &   24.11 &   23.42 &   23.94 &   22.00 &   20.36 &   
        1.48$^{+0.11}_{-0.09}$ \\
03d$^{\star}$ & 12:05:08.34 & -07:48:16.7 &  6.1 & $>$27.4 & $>$26.5 &   $\approx$25.3 &   $\approx$25.0 & $>$26.4 & $>$24.0 & $>$22.0 &   
        3.86$^{+0.56}_{-0.39}$ \\
03e & 12:05:07.89 & -07:48:05.9 &  6.7 &   24.75 &   23.90 &   22.82 &   22.09 &   22.61 &   20.78 &   19.27 &   
        0.41$^{+0.03}_{-0.01}$ \\
26a & 12:05:08.48 & -07:43:02.7 &  4.0 &   $\approx$26.7 &   !26.3 &   !25.3 &   !24.8 &   $\approx$25.6 & $>$24.0 & $>$22.0 &   
        0.56$^{+0.13}_{-0.17}$ \\
26b & 12:05:08.29 & -07:43:05.9 &  5.5 &   !26.9 &   !25.6 &   !25.8 &   $\approx$24.8 &   $\approx$26.2 & $>$24.0 & $>$22.0 &   
        2.91$^{+0.31}_{-0.23}$ \\
26c & 12:05:08.30 & -07:43:07.5 &  5.7 &   !27.1 & $>$26.5 &   25.04 &   23.88 &   24.80 &   23.30 & $>$22.0 &   
        0.67$^{+0.07}_{-0.11}$ \\
26d & 12:05:08.99 & -07:43:00.2 &  7.3 &   25.22 &   24.97 &   24.22 &   $\approx$24.6 &   24.50 &   $\approx$22.6 &   20.96 &   
        2.29$^{+0.04}_{-0.07}$ \\
26e & 12:05:08.51 & -07:42:58.6 &  7.4 &   25.62 &   $\approx$25.5 &   !25.4 & $>$25.5 &   $\approx$25.7 & $>$24.0 & $>$22.0 &   
        1.28$^{+1.55}_{-0.71}$ \\
34a & 12:05:09.51 & -07:40:04.5 &  4.8 &   22.88 &   22.69 &   21.83 &   21.45 &   22.28 &   20.73 &   19.56 &   
        0.51$^{+0.02}_{-0.01}$ \\
34b & 12:05:09.69 & -07:40:01.5 &  5.3 &   23.48 &   22.92 &   21.88 &   21.25 &   21.77 &   20.13 &   18.65 &   
        0.46$^{+0.02}_{-0.01}$ \\
34c & 12:05:10.16 & -07:40:04.0 &  5.9 &   22.76 &   21.60 &   20.59 &   19.82 &   20.42 &   18.59 &   17.14 &   
        0.40$^{+0.01}_{-0.01}$ \\
34d & 12:05:09.29 & -07:40:07.0 &  7.6 &   24.92 &   24.59 &   23.56 &   23.02 &   23.63 &   22.19 &   21.03 &   
        0.53$^{+0.03}_{-0.04}$ \\
34e & 12:05:10.33 & -07:40:08.5 &  8.1 &   25.50 &   $\approx$25.5 &   24.60 &   23.77 &   24.24 &   $\approx$22.6 &   21.08 &   
        1.27$^{+0.06}_{-0.07}$ \\
10a1 & 12:05:10.95 & -07:47:00.0 &  2.6 &   24.43 &   24.26 &   23.40 &   23.08 &   23.48 &   22.04 &   20.24 &   
        0.54$^{+0.03}_{-0.04}$ & 0.130\\
{\bf 10a2} & 12:05:10.85 & -07:47:00.0 & 2.4 & !25.9$^{\circ}$ & $\approx$25.55$^{\circ}$ & $\approx$25.0$^{\circ}$ & !24.8$^{\circ}$ & !25.6$^{\circ}$ & $>$24.0 & 20.58 & 
        2.49$^{+0.01}_{-0.05}$ & 0.136\\
10b & 12:05:10.94 & -07:46:54.0 &  3.6 &   24.41 &   23.92 &   23.29 &   22.56 &   23.04 &   20.65 &   19.17 &   
        2.09$^{+0.06}_{-0.05}$ \\
10c & 12:05:11.06 & -07:46:54.5 &  4.1 &   22.96 &   22.42 &   21.72 &   21.37 &   22.08 &   20.65 &   20.17 &   
        0.41$^{+0.01}_{-0.01}$ \\
10d & 12:05:10.65 & -07:46:54.8 &  4.3 &   25.44 &   25.13 &   24.15 &   23.21 &   23.96 &   22.16 &   20.94 &   
        0.60$^{+0.09}_{-0.07}$ \\
10e & 12:05:10.51 & -07:47:00.0 &  6.0 &   25.66 &   $\approx$25.9 &   25.08 &   $\approx$24.6 &   $\approx$25.8 &   !23.5 &   !21.9 &   
        0.55$^{+0.16}_{-0.11}$ \\
{\bf 36a} & 12:05:16.05 & -07:39:25.2 &  0.4 &   !27.1 & $>$26.5 & $>$26.2 &   $\approx$24.9 &   25.71 &   22.33 &   20.11 &   
        1.70$^{+0.12}_{-0.20}$ & 0.006\\
36b & 12:05:15.87 & -07:39:27.3 &  3.4 &   25.31 &   25.14 &   24.98 &   24.17 &   25.06 &   22.85 &   $\approx$21.4 &   
        1.55$^{+0.24}_{-0.12}$ \\
36c & 12:05:16.07 & -07:39:21.0 &  4.6 &   26.40 & $>$26.5 & $>$26.2 & $>$25.5 & $>$26.4 & $>$24.0 & $>$22.0 &   
        0.78$^{+1.90}_{-0.78}$ \\
36d & 12:05:15.62 & -07:39:24.8 &  6.6 & $>$27.4 & $>$26.5 &   $\approx$25.3 & $>$25.5 &   $\approx$25.9 & $>$24.0 & $>$22.0 &   
        4.04$^{+0.38}_{-0.63}$ \\
36e & 12:05:16.54 & -07:39:26.5 &  7.2 &   24.65 &   24.63 &   23.72 &   23.11 &   23.40 &   21.41 &   20.19 &   
        1.40$^{+0.01}_{-0.02}$ \\
{\bf 25a} & 12:05:17.89 & -07:43:08.6 &  0.4 &   $\approx$26.1 &   $\approx$25.3 &   $\approx$25.4 & $>$25.5 &   25.87 & $>$24.0 & 22.5$^{\bullet}$ &   
        2.91$^{+0.12}_{-0.14}$ & 0.020\\
25b & 12:05:18.24 & -07:43:08.5 &  5.7 &   25.26 &   25.14 &   24.73 &   24.21 &   24.61 &   22.79 & $>$22.0 &   
        1.41$^{+0.07}_{-0.11}$ \\
25c & 12:05:17.36 & -07:43:11.2 &  7.8 &   26.22 &   25.90 &   25.02 &   25.11 &   25.90 & $>$24.0 & $>$22.0 &   
        3.60$^{+0.08}_{-0.17}$ \\
25d & 12:05:18.40 & -07:43:09.3 &  8.1 &   25.04 &   24.52 &   23.42 &   22.41 &   22.80 &   21.02 &   19.38 &   
        0.61$^{+0.03}_{-0.03}$ \\

25e & 12:05:18.24 & -07:43:14.7 &  8.4 &   25.55 &   25.44 &   24.70 &   23.97 &   24.61 &   22.30 &   20.73 &   
        2.01$^{+0.25}_{-0.59}$ \\
01b & 12:05:20.08 & -07:49:33.6 &  3.4 &   25.59$^{\dagger}$ &   $\approx$25.5$^{\dagger}$ &   24.49$^{\dagger}$ &   24.63$^{\dagger}$ &   24.96$^{\dagger}$ &   23.28 &   20.31 &   
        0.50$^{+0.04}_{-0.05}$ \\
01c & 12:05:19.74 & -07:49:32.0 &  4.2 &   24.84 &   23.93 &   23.25 &   22.88 &   23.46 &   21.93 &   20.35 &   
        0.35$^{+0.01}_{-0.01}$ \\
{\bf 39a} & 12:05:20.59 & -07:38:55.4 &  0.6 &   25.43 &   25.14 &   24.17 &   22.73 &   23.19 &   20.99 &   19.04 &   
        1.32$^{+0.04}_{-0.01}$ & 0.005\\
39d & 12:05:20.22 & -07:38:55.0 &  6.2 &   26.24 & $>$26.5 &   25.17 & $>$25.5 &   25.70 &   $\approx$23.3 & $>$22.0 &   
        2.23$^{+0.16}_{-0.25}$ \\
39e & 12:05:20.66 & -07:39:02.6 &  7.1 &   22.75 &   22.07 &   21.26 &   20.74 &   21.50 &   19.77 &   19.20 &   
        0.36$^{+0.05}_{-0.02}$ \\
39f & 12:05:20.24 & -07:38:49.9 &  8.1 &   24.76 &   24.54 &   23.49 &   22.12 &   22.43 &   20.14 &   18.35 &
        1.37$^{+0.02}_{-0.01}$ \\
{\bf 18a} & 12:05:23.13 & -07:45:14.9 &  0.3 &   $\approx$26.9 & $>$26.5 &   24.87 &   23.34 &   23.59 &   21.56 &   19.61 &   
        1.25$^{+0.06}_{-0.06}$ & 0.003\\
18b & 12:05:23.30 & -07:45:15.8 &  2.9 &   !27.1 &  $\approx$26.2 &   $\approx$25.8 &   !25.3 &   $\approx$25.6 & $>$24.0 & $>$22.0 &   
        3.10$^{+0.40}_{-0.54}$ \\
18c & 12:05:22.93 & -07:45:13.0 &  3.2 &   25.75 &   $\approx$26.3 &   $\approx$26.1 &   24.37 &   25.38 &   $\approx$23.4 &  21.40 &   
        1.29$^{+0.10}_{-0.25}$ \\
18d & 12:05:22.88 & -07:45:16.8 &  4.3 &   25.29 &   24.55 &   23.55 &   22.60 &   22.85 &   20.64 &   18.69 &   
        2.40$^{+0.03}_{-0.02}$ \\
18e & 12:05:23.36 & -07:45:11.5 &  4.6 &   $\approx$26.9 & $>$26.5 &   $\approx$25.1 & $>$25.5 &   $\approx$26.2 & $>$24.0 & $>$22.0 &   
        3.77$^{+0.12}_{-0.23}$ \\
08a & 12:05:24.65 & -07:47:20.5 &  3.8 &   $\approx$27.2 & $>$26.5 &   $\approx$25.8 &   $\approx$25.0 &   $\approx$25.6 & $>$24.0 & $>$22.0 &   
        0.66$^{+0.20}_{-0.29}$ \\
08b & 12:05:25.06 & -07:47:19.5 &  5.6 & $>$27.4 & $>$26.5 & $>$26.2 &   $\approx$24.8 &   25.58 &   !23.8 & $>$22.0 &   
        5.25$^{+0.40}_{-0.76}$ \\
08c & 12:05:24.94 & -07:47:17.6 &  6.3 &   25.18 &   24.66 &   23.91 &   22.89 &   23.45 &   21.81 &   20.48 &   
        0.74$^{+0.04}_{-0.06}$ \\
08d & 12:05:24.30 & -07:47:22.5 &  7.6 &   26.59 & $>$26.5 & $>$26.2 & $>$25.5 &   !26.0 & $>$24.0 & $>$22.0 &   
        1.23$^{+0.29}_{-0.70}$ \\
08e & 12:05:24.75 & -07:47:15.9 &  7.7 &   26.09 & $\approx$26.1 & $>$26.2 &   $>$25.5 & $>$26.4 & $>$24.0 & $>$22.0 &   
        0.91$^{+1.75}_{-0.91}$ \\
{\bf 13a} & 12:05:26.78 & -07:46:41.0 &  0.5 &   $\approx$26.5 & $>$26.5 &   !25.6 &  !24.9 &   !25.7 &   $\approx$23.1 &   20.44 &   
        2.26$^{+0.21}_{-0.59}$ & 0.014\\
13c & 12:05:26.96 & -07:46:45.3 &  5.6 &   $\approx$26.8 & $>$26.5 &   25.30 &   24.84 &   25.63 &   !23.4 & $>$22.0 &   
        0.50$^{+0.12}_{-0.50}$ \\
13d & 12:05:26.78 & -07:46:34.9 &  5.7 &  24.87 &   24.52 &   23.91 &   23.51 &   23.80 &   22.33 &   21.40 &   
        1.36$^{+0.08}_{-0.04}$ \\
13e & 12:05:26.29 & -07:46:39.4 &  7.1 &  25.17 &   24.90 &   23.66 &   23.11 &   23.77 &   22.21 &   20.99 &   
        0.53$^{+0.03}_{-0.02}$ \\
{\bf 29a}& 12:05:30.26 & -07:41:45.4 &  0.2&   25.48 &   25.17 &   24.46 &   24.01 &   25.06 &   22.60 &   20.68 &   
        2.26$^{+0.10}_{-0.10}$ & 0.003\\
29b & 12:05:30.17 & -07:41:46.8 &  1.8 &   25.33 &   24.84 &   24.44 &   24.44 &   25.25 &   22.88 &   21.40 &   
        2.34$^{+0.07}_{-0.09}$ \\
\tablebreak
\tablebreak
\tablebreak
\tablebreak
\tablebreak
\tablebreak
\tablebreak
\tablebreak
\tablebreak
\tablebreak
\tablebreak
29c & 12:05:30.13 & -07:41:44.2 &  2.3 &   !27.3 & $>$26.5 &   !25.6 & $>$25.5 & $>$26.4 & $>$24.0 &   21.23 &   
        2.70$^{+0.48}_{-0.34}$ \\
29d & 12:05:29.91 & -07:41:46.4 &  5.2 &  26.11 &  $\approx$26.0 & $>$26.2 & $>$25.5 &   26.01 & $>$24.0 & $>$22.0 &   
        1.23$^{+1.69}_{-1.23}$ \\
29e & 12:05:30.00 & -07:41:51.2 &  6.9&   26.59 &   $\approx$26.1 &   $\approx$25.8 &   24.80 &   25.68 &   $\approx$23.1 & $>$22.0 &   
        1.72$^{+0.34}_{-0.21}$ \\
07a & 12:05:31.01 & -07:47:39.9 &  0.9&   26.06 &   $\approx$25.6 &   $\approx$25.2 & $>$25.5 &   25.34 & $>$24.0 & $>$22.0 &   
        0.35$^{+0.34}_{-0.35}$ & 0.108\\
07b & 12:05:31.19 & -07:47:37.1 &  3.1& $>$27.4 & $>$26.5 &   25.33 &   !25.2 &   25.61 &   $\approx$23.3 &   $\approx$21.9 &   
        2.49$^{+0.17}_{-0.18}$ \\
07c & 12:05:30.87 & -07:47:44.2 &  5.5&   $\approx$26.4 &   !26.0 &   24.66 &   23.64 &   24.25 &   22.14 &   21.53 &   
        0.47$^{+0.09}_{-0.47}$ \\
07d& 12:05:31.30 & -07:47:34.7 &  6.0&   26.23 &   25.29 &   24.81 &   !24.9 &   26.05 & $>$24.0 & $>$22.0 &   
        3.67$^{+0.06}_{-0.09}$ \\
07e& 12:05:30.64 & -07:47:39.9 &  6.3&   23.39 &   22.81 &   22.12 &   21.74 &   22.48 &   21.04 &   20.51 &   
        0.41$^{+0.01}_{-0.01}$ \\
04a & 12:05:31.17 & -07:48:04.7 &  1.0&   22.39$^{\ast}$ &   21.64$^{\ast}$ &   21.14$^{\ast}$ &   20.27$^{\ast}$ &   20.89 &   19.02 &   17.36 &   
        0.25$^{+0.03}_{-0.04}$ & 0.005\\
04b & 12:05:31.01 & -07:48:07.1 &  2.3&   23.43$^{\ast}$ &   23.32$^{\ast}$ &   23.23$^{\ast}$ &   22.31 &   23.09 &   21.14 &   19.51 &   
        1.39$^{+0.06}_{-0.01}$ \\
04d & 12:05:31.21 & -07:48:11.0 &  5.5&  25.97 &   !26.1 & $>$26.2 &   $\approx$25.3 &   25.41 &   22.30 &   20.33 &   
        1.83$^{+0.01}_{-0.05}$ \\
04e & 12:05:31.47 & -07:48:08.6 &  5.9&  23.52 &   22.80 &   21.81 &   21.10 &   21.68 &   20.08 &   18.57 &   
        0.41$^{+0.05}_{-0.02}$ \\
{\bf 42a} & 12:05:34.93 & -07:38:17.4 &  1.3 & $>$27.4 & $>$26.5 & $>$26.2 &   $\approx$24.9 &   !25.7 &   !22.9 &   20.94 &   
        1.52$^{+0.24}_{-0.34}$ & 0.060\\
42b & 12:05:34.96 & -07:38:19.7 &  1.5 &   26.28 &   $\approx$25.7 &   !25.4 &   !25.3 & $>$26.4 & $>$24.0 &   !21.5 &   
        2.83$^{+0.67}_{-0.24}$ & 0.082\\
42c & 12:05:34.77 & -07:38:17.7 &  2.0 & $>$27.4 &   26.24 & $>$26.2 & $>$25.5 &   !25.5 & $>$24.0 &   21.33 &   
        3.10$^{+0.34}_{-0.44}$ \\
42d & 12:05:35.06 & -07:38:22.0 &  4.2 & $>$27.4 & $>$26.5 & $>$26.2 &   !25.4 & $>$26.4 & $>$24.0 &   21.04 &   
        4.21$^{+0.79}_{-1.67}$ \\
16a & 12:05:39.61 & -07:45:27.9 &  2.3 &   25.54 &   25.30 &   24.85 &   23.63 &   24.32 &   $\approx$23.8 &   21.11 &   
        0.93$^{+0.13}_{-0.08}$ \\
16b & 12:05:39.35 & -07:45:28.7 &  2.4 &   25.47 &   25.27 &   24.78 &   23.88 &   24.95 &   22.21 &   20.50 &   
        1.70$^{+0.19}_{-0.12}$ \\
16c & 12:05:39.54 & -07:45:24.3 &  2.9 &   25.20 &   24.58 &   23.59 &   23.06 &   23.59 &   21.97 &   20.45 &   
        0.46$^{+0.02}_{-0.02}$ \\
16d & 12:05:39.19 & -07:45:22.7 &  6.0 &   26.21 &   $\approx$26.3 &   $\approx$26.0 &   24.62 &   25.16 &   $\approx$23.5 &   $\approx$21.9 &   
        1.13$^{+0.26}_{-0.17}$ \\
16e & 12:05:39.93 & -07:45:25.0 &  7.1 &   26.05 &   !26.4 &   $\approx$25.4 &   $\approx$25.0 &   25.65 & $>$24.0 & $>$22.0 &   
        0.58$^{+0.18}_{-0.10}$ \\
40a & 12:05:45.69 & -07:38:50.5 &  2.1 &   26.33 &   $\approx$26.1 &   $\approx$25.9 &   !25.0 &   !26.1 &  !23.7 &   !21.7 &   
        1.61$^{+0.92}_{-0.25}$ & 0.168\\
{\bf 40b} & 12:05:45.98 & -07:38:51.8 &  3.9 & $>$27.4 & $>$26.5 & $>$26.2 &   !25.3 &   $\approx$25.8 &   22.19 &   19.57 &   
        1.52$^{+1.24}_{-0.25}$ \\
40c & 12:05:45.55 & -07:38:48.4 &  4.8 &   $\approx$26.7 & $>$26.5 &   $\approx$25.2 &   !25.2 &   25.40 & $>$24.0 & $>$22.0 &   
        0.54$^{+0.09}_{-0.10}$ \\
40d & 12:05:45.69 & -07:38:59.1 &  6.6 &   25.24 &   24.63 &   23.92 &   23.60 &   23.99 &   !23.2 &   21.41 &   
        3.41$^{+0.05}_{-0.08}$ \\
31a & 12:05:46.69 & -07:41:30.3 &  4.3 &   25.80 &   25.04 &   $\approx$24.8 &   $\approx$23.8 &   25.12 &  !23.6 &   $\approx$21.8 &   
        2.87$^{+0.11}_{-0.17}$ \\
31b & 12:05:46.94 & -07:41:33.3 &  5.3 &   25.52 &   25.70 &   25.33 &   $\approx$24.7 &   25.08 &   $\approx$23.1 & $>$22.0 &   
        1.35$^{+0.15}_{-0.08}$ \\
31c & 12:05:46.65 & -07:41:39.5 &  5.3 &   25.72 &   25.23 &   24.48 &   23.30 &   23.75 &   22.29 &   20.37 &   
        1.11$^{+0.05}_{-0.03}$ \\
31d & 12:05:46.82 & -07:41:38.9 &  5.7 &   $\approx$26.8 & $>$26.5 &   25.01 &   23.82 &   24.58 &   $\approx$23.0 & $>$22.0 &   
        0.70$^{+0.09}_{-0.08}$ \\
\tablecomments{
Col. (1) --- Optical/near-infrared source identification. Suggested
counterparts to mm sources are marked in bold.\\
Col. (2)-(3) --- J2000 coordinates of optical/near-infrared source.\\
Col. (4) --- Separation between optical position and best interferometric 
position.\\
Col. (5)-(11) --- Source magnitudes in a 2\arcsec\ aperture. All magnitudes 
are on the Vega system except for the z magnitudes which are on the BD 
system (see text).\\ 
Col. (12) --- Optical/near-infrared photometric redshift.\\
Col. (13) --- Probability of a chance coincidence with an object of the given K
magnitude (07a: B) at the given separation.\\
$\dagger$: magnitude contaminated by nearby foreground galaxy.\\
$\ast$: magnitude contaminated by nearby foreground star.\\
$\star$: 03d is not detected in B,z and K but retained due to credible
R and I detections.\\
$\circ$: magnitude contaminated by 10a1.\\
$\bullet$: Lehnert et al. (2004, in preparation). 
}
\enddata
\label{tab:idtable}
\end{deluxetable}

\clearpage
\begin{deluxetable}{rrrrrrrrr}
\tabletypesize{\tiny}
\tablecolumns{9}
\tablewidth{0pt}
\tablecaption{Main properties of counterparts to MAMBO sources}
\tablehead{

\colhead{Source}&\colhead{No}&\colhead{Q}&\colhead{S$_{1.2mm}$}
&\colhead{S$_{1.4}$}&\colhead{K}&\colhead{z$_{CY}$}&\colhead{z$_{phot}$}
&\colhead{z$_{submm/mm}$}\\
\colhead{}&\colhead{}&\colhead{}&\colhead{mJy}
&\colhead{$\mu$Jy}&\colhead{mag}&\colhead{}&\colhead{}
&\colhead{}\\
\colhead{(1)}&\colhead{(2)}&\colhead{(3)}&\colhead{(4)}
&\colhead{(5)}&\colhead{(6)}&\colhead{(7)}&\colhead{(8)}
&\colhead{(9)}
}
\startdata
MMJ120507-0748.1 & 03&g&4.6& 88&   21.48&
2.38$^{+1.19}_{-0.76}$&2.11$^{+0.26}_{-0.61}$&4.25$^{+\infty}_{-4.25}$\\
MMJ120508-0743.1 & 26&u&2.7& 42&$>$22.0 &
2.64$^{+1.55}_{-0.95}$&    &    \\
MMJ120509-0740.0 & 34&u&3.3& 47&$>$22.0 &
2.73$^{+1.49}_{-0.91}$&    & (see note)\\
MMJ120510-0747.0 & 10&u&3.1& 41&   20.58&
2.82$^{+1.74}_{-1.00}$&2.49$^{+0.01}_{-0.05}$&    \\
MMJ120516-0739.4 & 36&g&2.2&459&   20.11&
0.83$^{+0.42}_{-0.34}$&1.70$^{+0.12}_{-0.20}$&    \\
MMJ120517-0743.1 & 25&g&4.3& 40&   22.50&
3.27$^{+2.07}_{-1.08}$&2.91$^{+0.12}_{-0.14}$&3.65$^{+4.26}_{-2.53}$\\
MMJ120519-0749.5 & 01&g&5.2& 72&$>$22.0 &
2.74$^{+1.41}_{-0.87}$&    &$>8.95$\\
MMJ120520-0738.9 & 39&g&2.7&336&   19.04&
1.05$^{+0.50}_{-0.39}$&1.32$^{+0.04}_{-0.01}$&     \\
MMJ120522-0745.1 & 18&g&2.0& 90&   19.61&
1.67$^{+0.79}_{-0.58}$&1.25$^{+0.06}_{-0.06}$&     \\
MMJ120524-0747.3 & 08&u&2.0& 47&$>$22.0 &
2.21$^{+1.25}_{-0.79}$&    &     \\
MMJ120526-0746.6 & 13&g&2.6& 41&   20.44&
2.62$^{+1.54}_{-0.92}$&2.26$^{+0.21}_{-0.59}$&     \\
MMJ120530-0741.6 & 29&g&2.3& 85&   20.68&
1.82$^{+0.88}_{-0.63}$&2.26$^{+0.10}_{-0.10}$&     \\
MMJ120530-0747.7 & 07&u&3.0& 40&$>$22.0 &
2.82$^{+1.77}_{-1.00}$&(0.35$^{+0.34}_{-0.35}$)&     \\
MMJ120531-0748.1 & 04&u&2.7& 53&$>$18.0 &
2.37$^{+1.32}_{-0.86}$&(0.25$^{+0.03}_{-0.04}$)&     \\
MMJ120534-0738.3 & 42&g&2.2& 61&   20.94&
2.05$^{+1.14}_{-0.79}$&1.52$^{+0.24}_{-0.34}$&     \\
MMJ120539-0745.4 & 16&g&3.4& 55&$>$22.7 &
2.56$^{+1.37}_{-0.84}$&    &6.35$^{+\infty}_{-4.85}$\\
MMJ120545-0738.8 & 40&u&3.7& 40&   19.57&
3.05$^{+1.97}_{-1.04}$&1.52$^{+1.24}_{-0.25}$&     \\
MMJ120546-0741.5 & 31&g&6.5& 42&   21.90&
3.86$^{+2.58}_{-1.31}$&    &5.25$^{+6.14}_{-1.93}$\\
\tablecomments{
Col. (1) --- MAMBO source.\\ 
Col. (2) --- Short number on the original MAMBO NDF source list.\\
Col. (3) --- Quality of the interferometric identification. g(ood) stands
  for objects with a good quality VLA source or a PdBI identification.\\
Col. (4) --- MAMBO 1.2mm peak flux density.\\
Col. (5) --- VLA 1.4GHz peak flux density.\\
Col. (6) --- K magnitude (Vega), see Table~\ref{tab:idtable} for other
bands. For MMJ120519-0749.5 and MMJ120531-0748.1, we give K-band limits as we
assume that the foreground objects detected at the radio position are not
linked to the dust emission. \\
Col. (7) --- Redshift estimate from radio/mm spectral index 
             \citep{carilli00b}. Errors reflect 1$\sigma$ uncertainties in 
             alpha due to measurement errors and due to the intrinsic scatter
             of the relation, added in quadrature.\\
Col. (8) --- Photometric redshift from optical/near-infrared photometry. 
   Values in brackets refer to objects - MMJ120519-0749.5 and MMJ120531-0748.1
   - that are likely in the foreground.\\
Col. (9) --- Redshift estimate from the 1200$\mu$m/850$\mu$m flux ratio as
             derived by \citet{eales03}. No. 34 was observed but the flux 
             ratio could not be fit by the templates used at any redshift.
}
\enddata
\label{tab:counter}
\end{deluxetable}
%
% figures
%
\clearpage
\figcaption[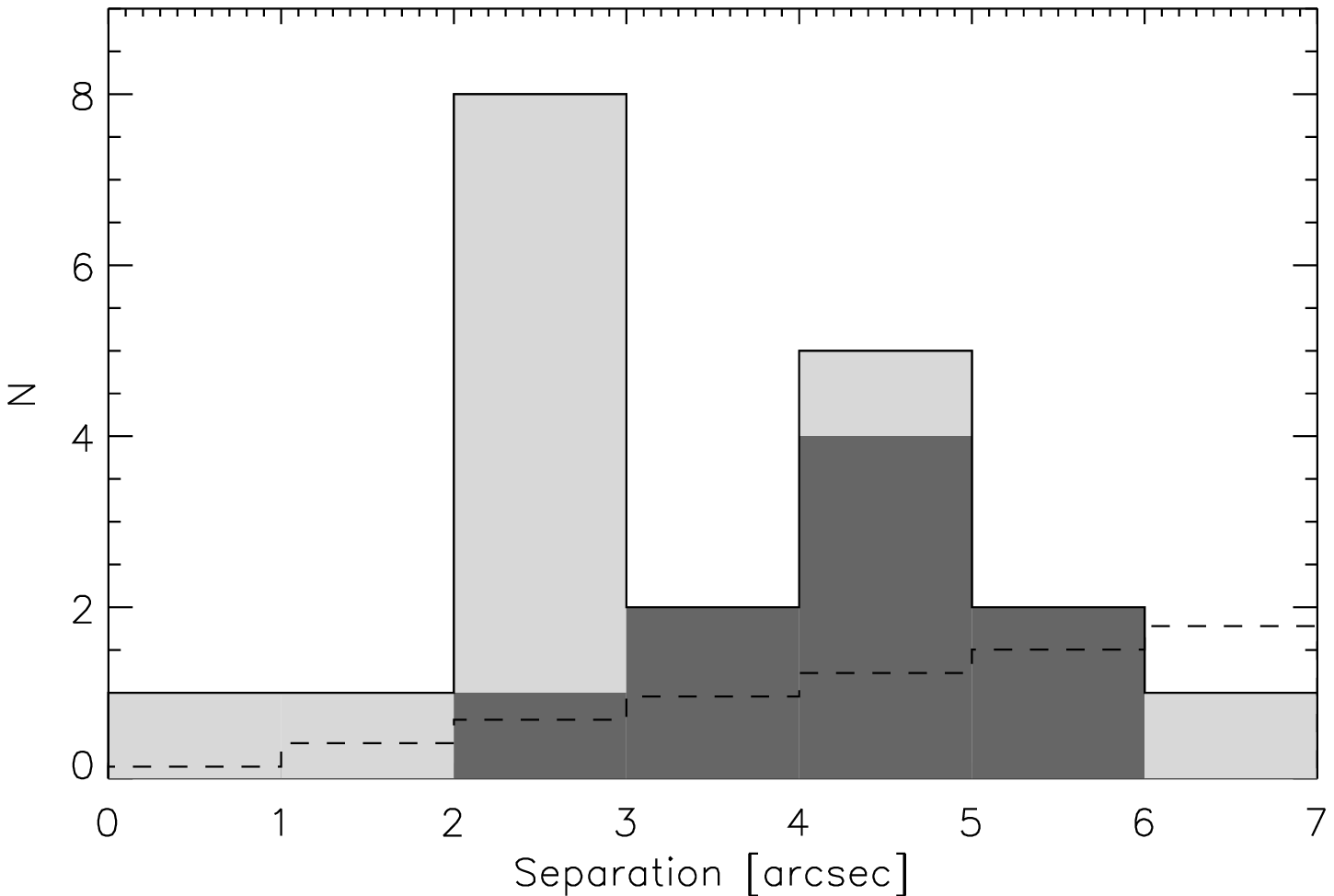]{Histogram of separations between mm positions
from the MAMBO map and VLA positions of associated radio sources. Light and
dark grey indicate the good and uncertain sources from Table~\ref{tab:assoc}.
The dashed histogram indicates the expected number of unrelated background
radio peaks.  
\label{fig:assoc}}
\figcaption[f2.eps]{BRzK images (from left to right) for the fields
of the MAMBO sources with radio associations. The $25\arcsec\times
25\arcsec$ images are centered on the nominal MAMBO positions listed
in Table~\ref{tab:assoc} and oriented such that north is at the top
and east to the left. A small cross indicates the best interferometric
position (PdBI if available, VLA otherwise) and its 1$\sigma$ error. The
right panel indicates all $\geq 40\mu$Jy radio peaks by crosses, and the
7\arcsec\ radius used to search for radio peaks possibly associated with
the MAMBO sources.  Small letters label optical/near-infrared sources near
to the best interferometric position (see also Table~\ref{tab:idtable}).
\label{fig:idplot1}}
\figcaption[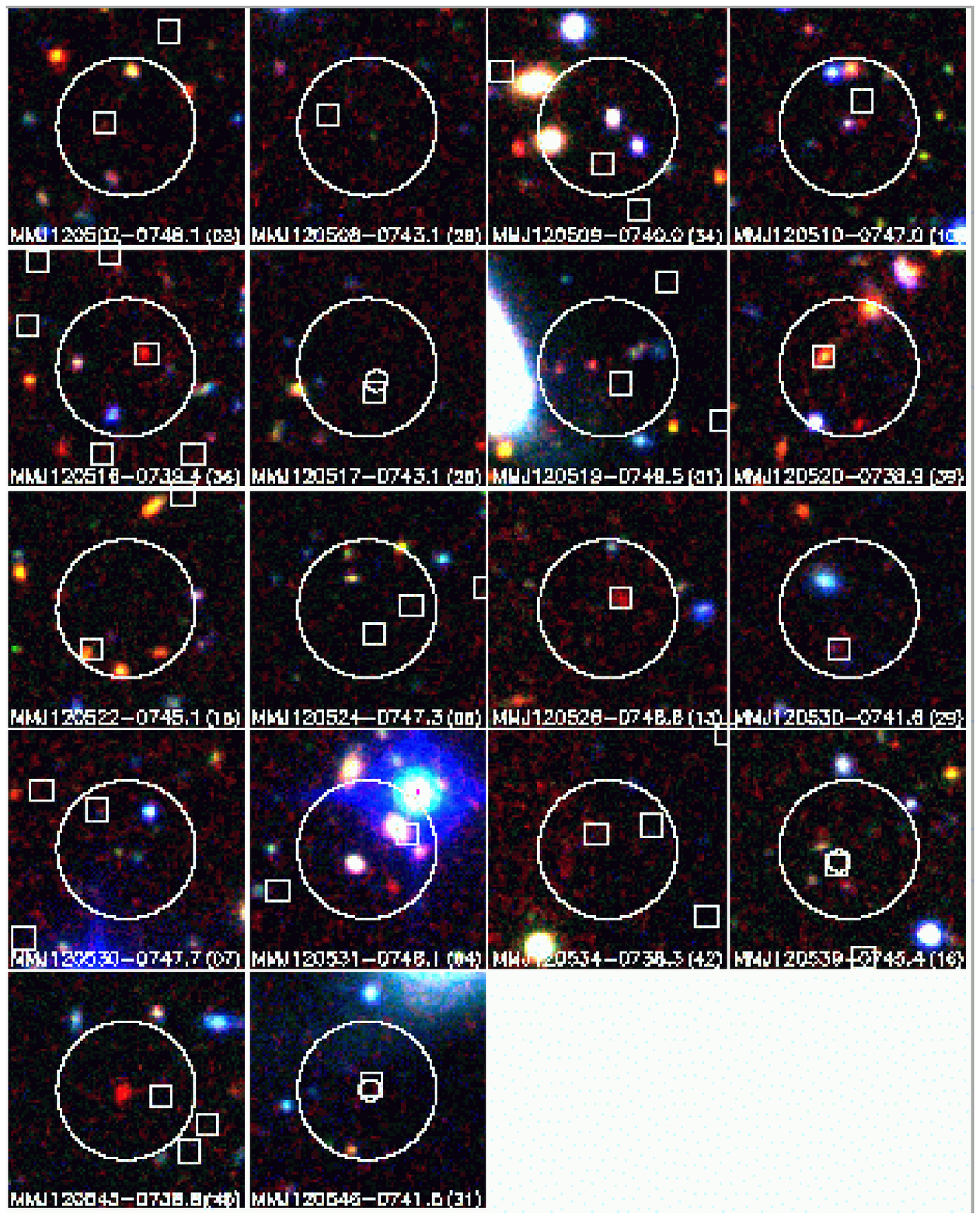]{BzK color composites of the NDF MAMBO sources with
interferometric positions. Large 7\arcsec\ radius circles are drawn
around the nominal MAMBO positions. Unlike in Fig. 2, we now show both
the VLA 1.4GHz peaks (indicated by squares) and position of the PdBI mm
interferometric sources (indicated by small circles).
\label{fig:colorcomp}}
\figcaption[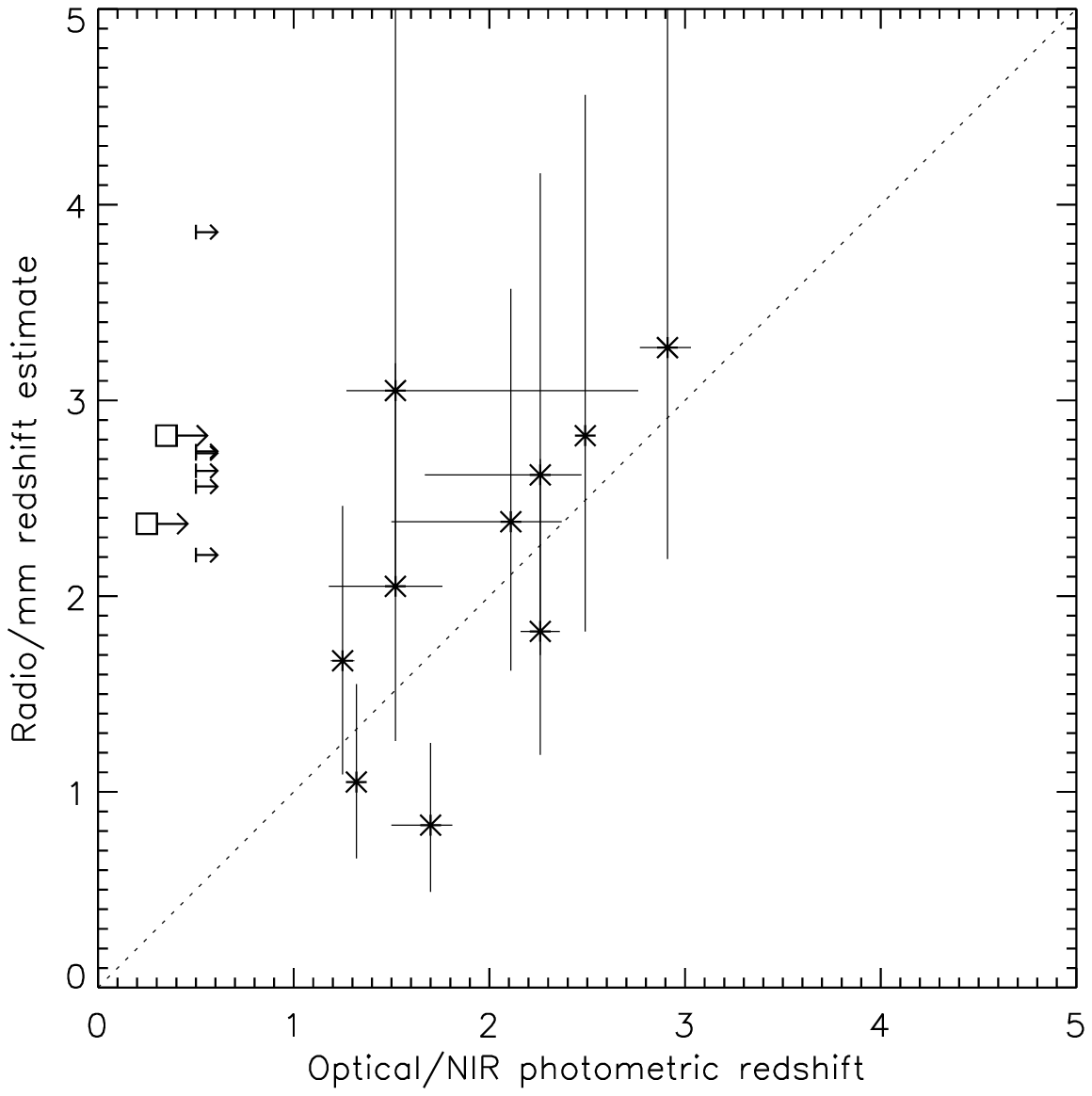]{Comparison of optical/near-infrared photometric
redshifts with redshift estimates from the radio/mm spectral index. Blank
fields without photometric redshifts are shown with an arbitrary lower
limit of 0.5 for the optical/near-infrared photometric redshift, and
without the error bars for the radio/mm estimate. For two objects shown
by squares we believe that the measured optical/NIR photometric redshift
refers to a foreground object, the true redshift of the submm source
being higher (see text).
\label{fig:photvscy}}
\figcaption[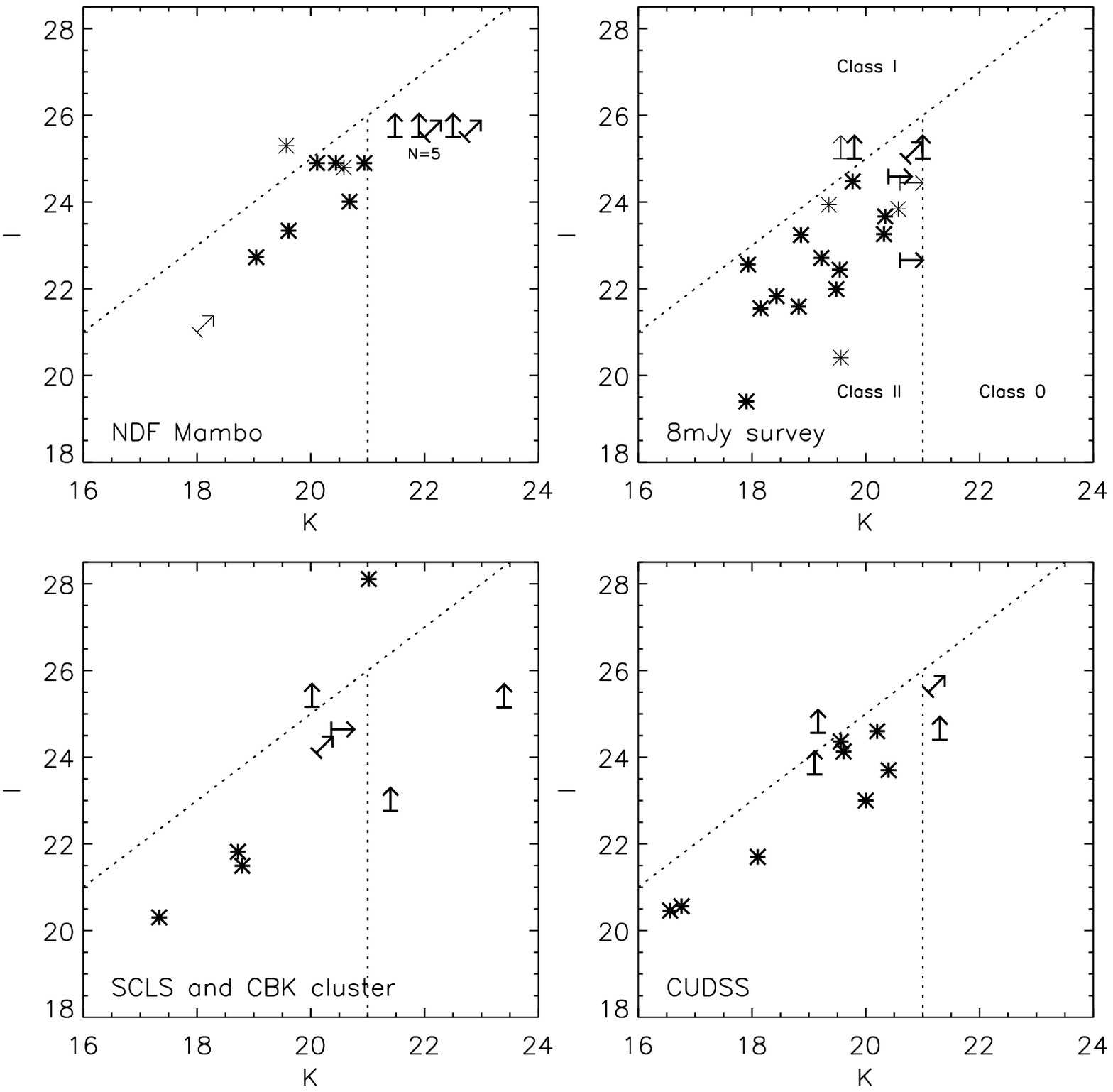]{Distribution of K and I magnitudes for the
interferometrically located NDF 1.2mm objects and for three samples
of interferometrically located SCUBA 850$\mu$m galaxies. The limit
symbol at K$>$22, I$>25.5$ in the NDF panel stands for 5 objects.
In the panel for the 8mJy survey, we indicate the regions occupied
in this type of diagram by the class 0/I/II objects in the definition
used by \cite{ivison02}. In the top two panels thick symbols reflect
identifications described as good/robust.  Magnitudes for the cluster
lens sources have been corrected for the magnifications stated in the
original papers.
\label{fig:magdist}}
\figcaption[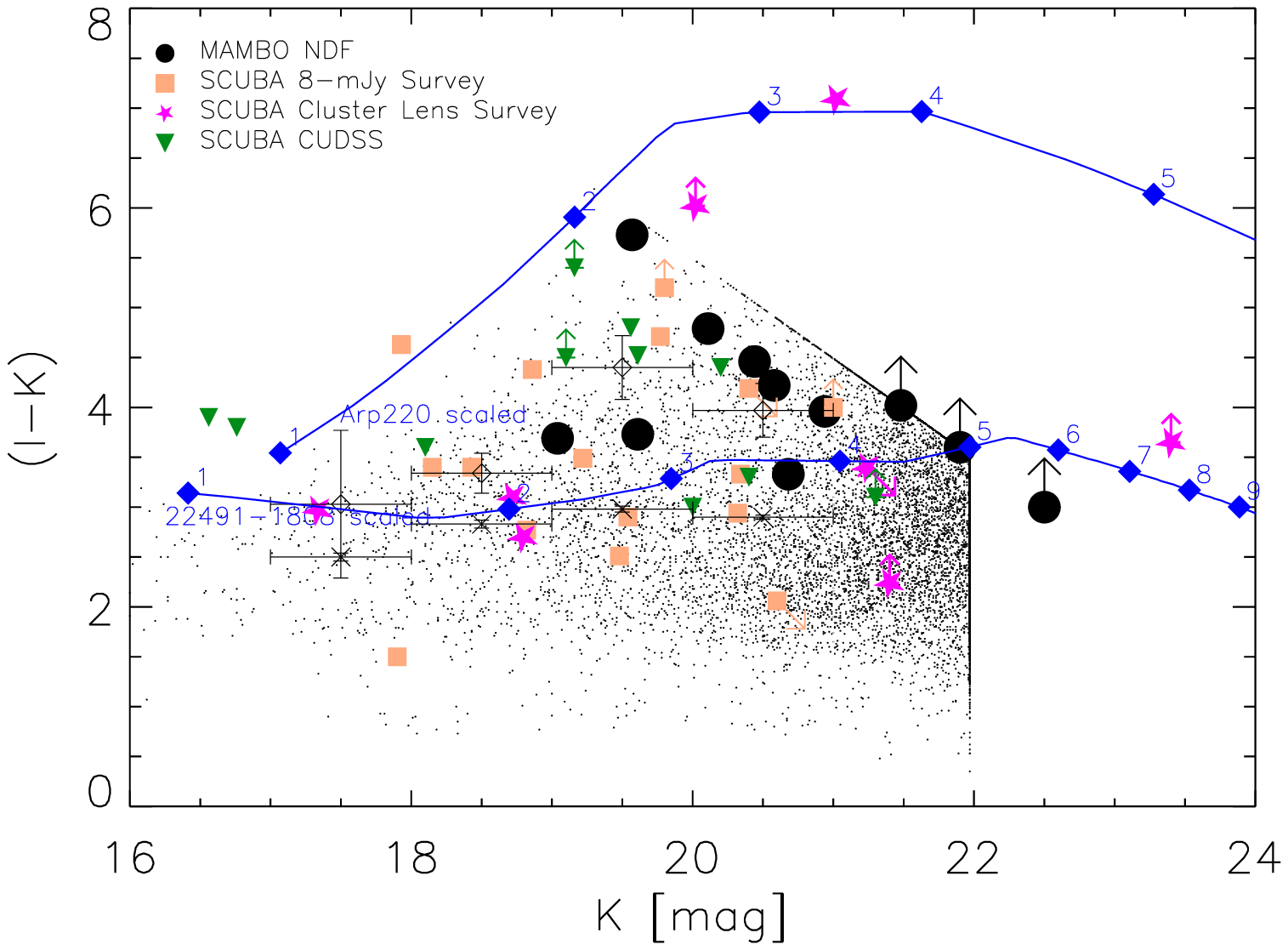]{
I-K vs. K magnitude-color diagram for the MAMBO identifications (large
filled circles). Small dots show the field galaxy population from our
NDF data. Interferometrically located SCUBA galaxies are represented
by squares for the 8mJy sample objects from \citet{ivison02}, stars
for cluster lens survey objects (corrected for magnification)
from \citet{smail02}, and triangles for the CUDSS objects of
\citet{webb03a,webb03b}. Filled diamonds connected by continuous lines
show the expected colors/magnitudes for redshift 1,2,\ldots for objects
with the SED  shapes of Arp220 and IRAS 22491-1808 {\em but scaled to
observed S$_{1.2mm}$=5mJy}. Average properties in magnitude bins are shown
with error bars for the (sub)mm population (open diamonds) including
all objects shown, and for the field population (crosses). Blank field
sources are not included, the figure is an incomplete representation in
its right part.
\label{fig:ik}}
\figcaption[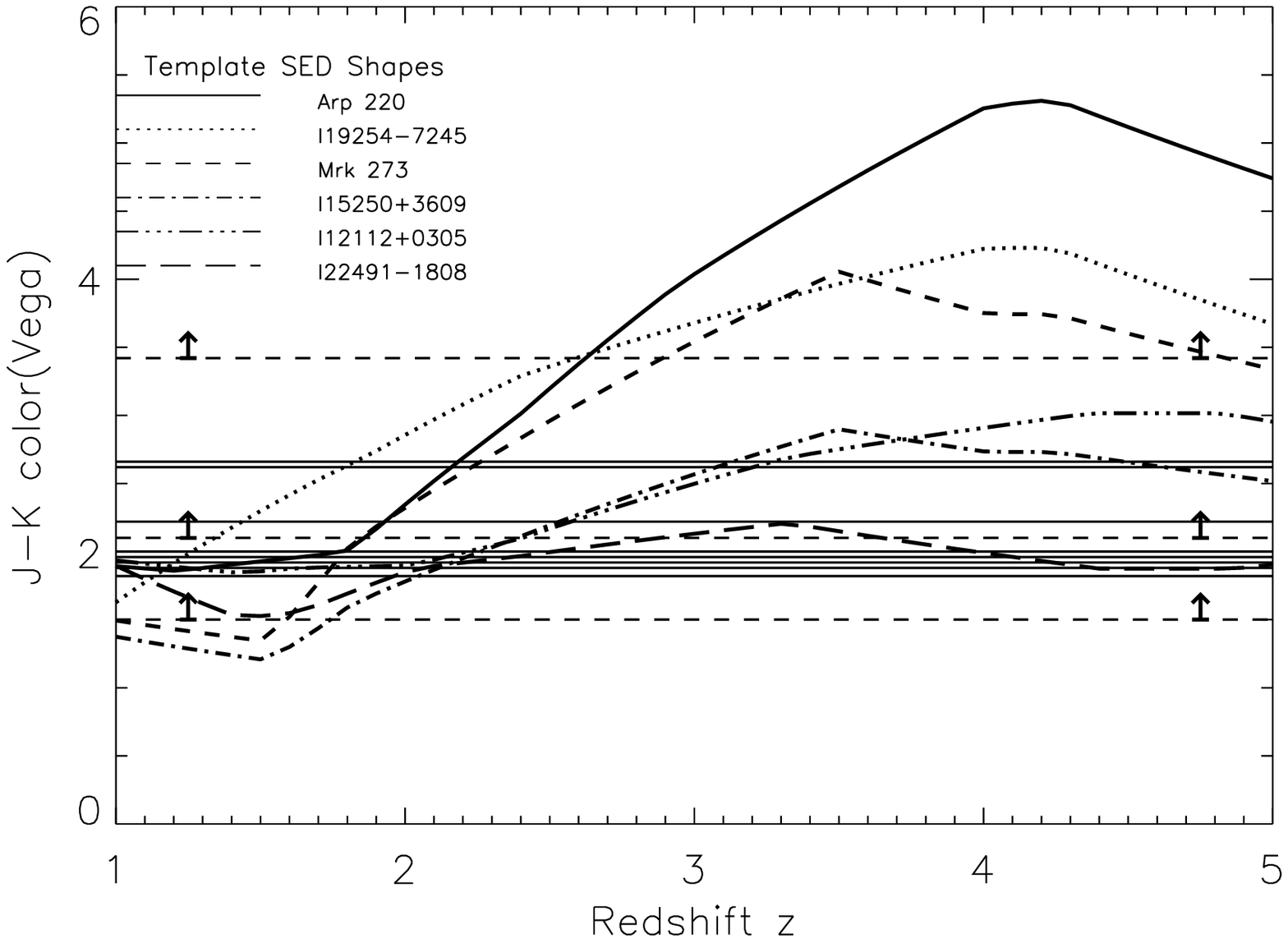]{J-K color as a function of redshift for the
template SED shapes of a number of local ultra-luminous infrared galaxies. 
Horizontal over-plotted lines show measurements (continuous) or lower
limits (dashed) for the J-K color our MAMBO sources. 
\label{fig:jk}}
\figcaption[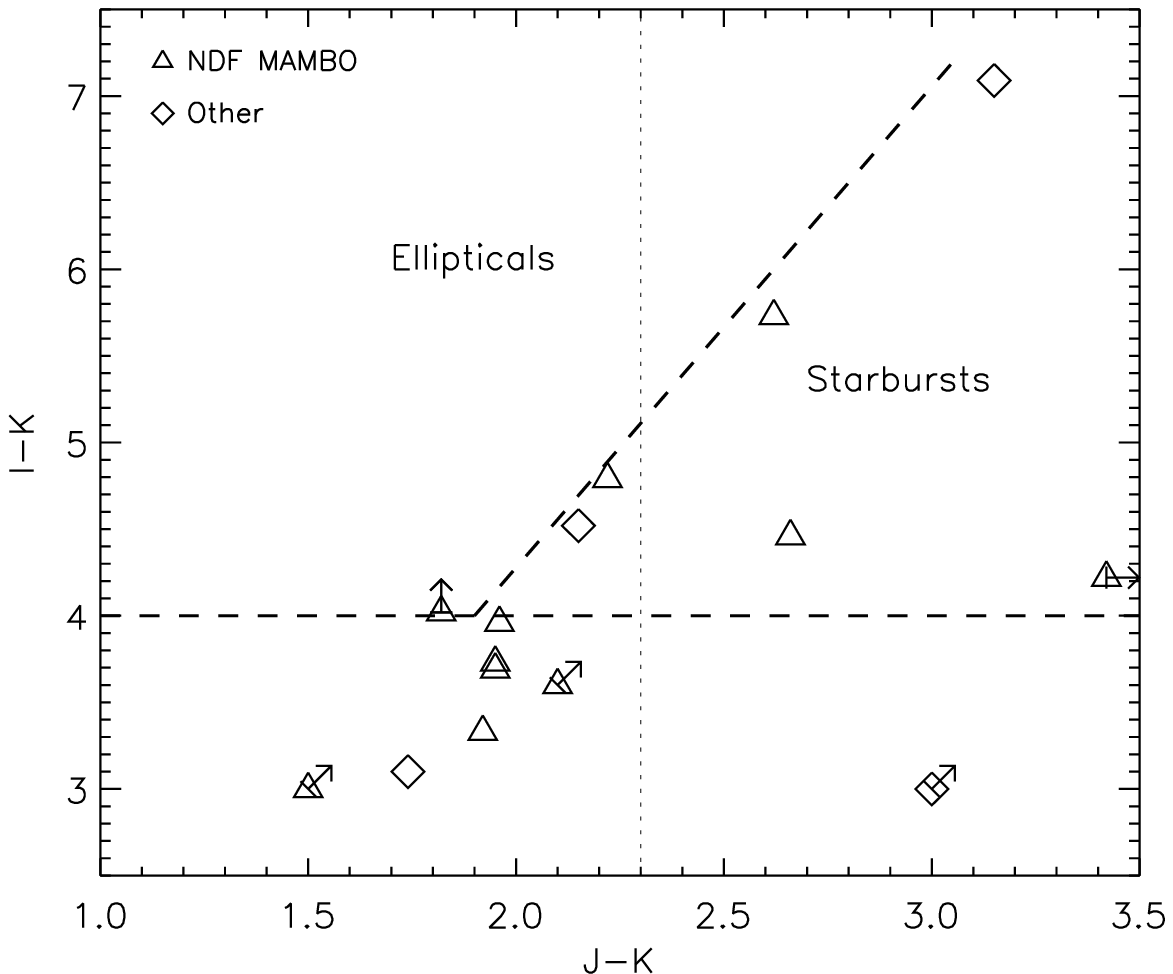]{I-K vs. J-K diagram, with the thick dashed lines
showing the regions populated by z$\sim$1-2 passive ellipticals and by
dusty starbursts according to \citet{pozzetti00}. Triangles indicate
NDF MAMBO sources. Diamonds indicate SCUBA sources from the literature with IJK
photometry and one A2125 MAMBO source from \citet{bertoldi00}. The thin 
vertical dashed line indicates the J-K$>$2.3 criterion of \citet{franx03}. 
\label{fig:pozzetti}}
\figcaption[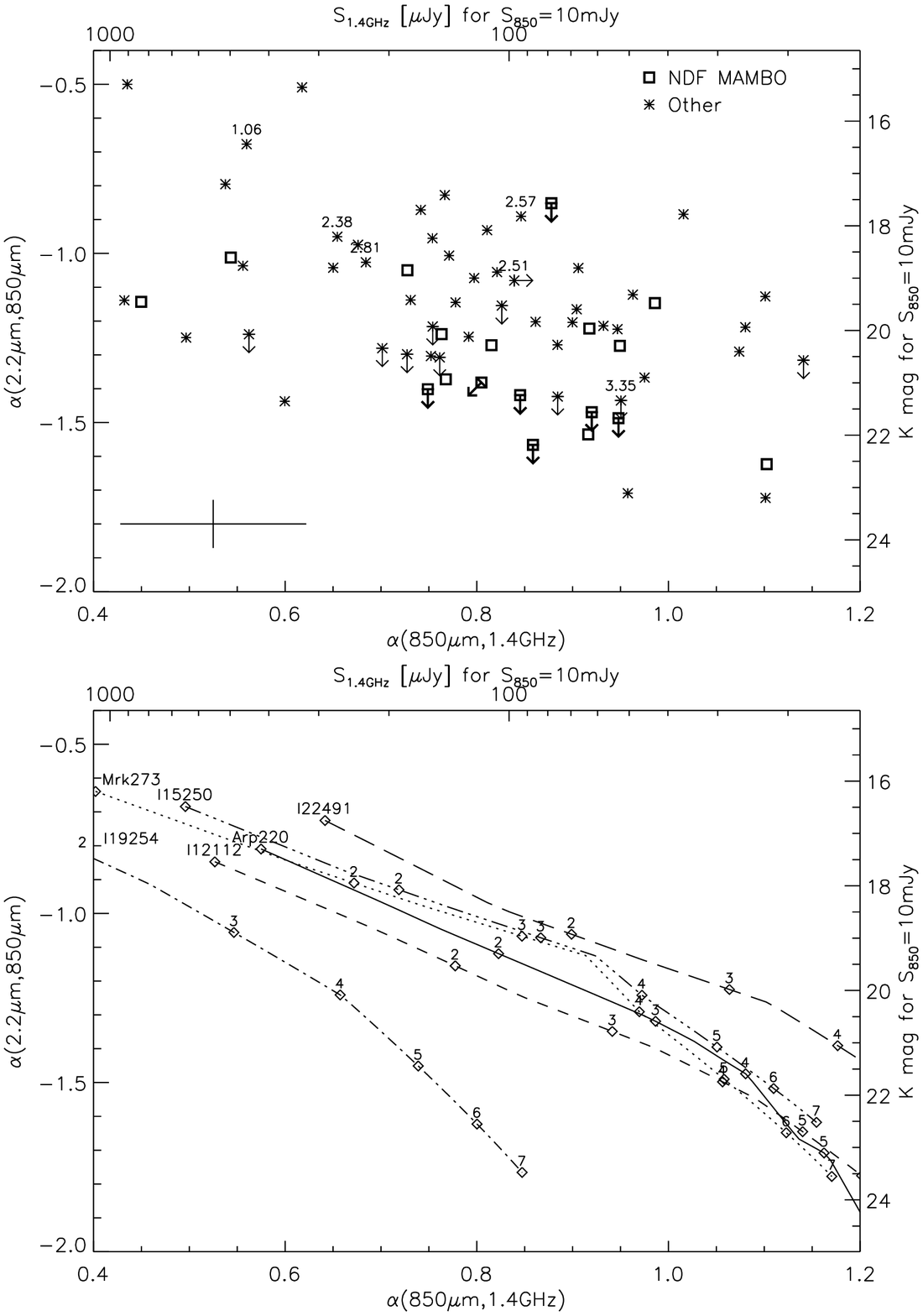]{Top panel: Diagnostic diagram for the near-infrared
to radio SEDs of (sub)mm sources, combining the radio/submm spectral
index with a spectral index between near-infrared and submm. Thick
squares/limits represent the NDF MAMBO source discussed in this paper,
thin asterisks/limits mostly SCUBA sources from the samples described
in the text, and a few individual well studied (sub)mm objects.
Objects with CO-confirmed redshifts are labeled with their redshift.
A typical observational error is indicated in the lower left.  The lower
panel shows for comparison the loci corresponding to the redshifted SEDs
of a number of local ULIRGs with available UV to radio SEDs. Note that
IRAS 19254-7245 hosts a powerful AGN.
\label{fig:rgdiag}}
\clearpage

\epsscale{0.6}
\plotone{f1.eps}
\newpage
\epsscale{1.0}
\plotone{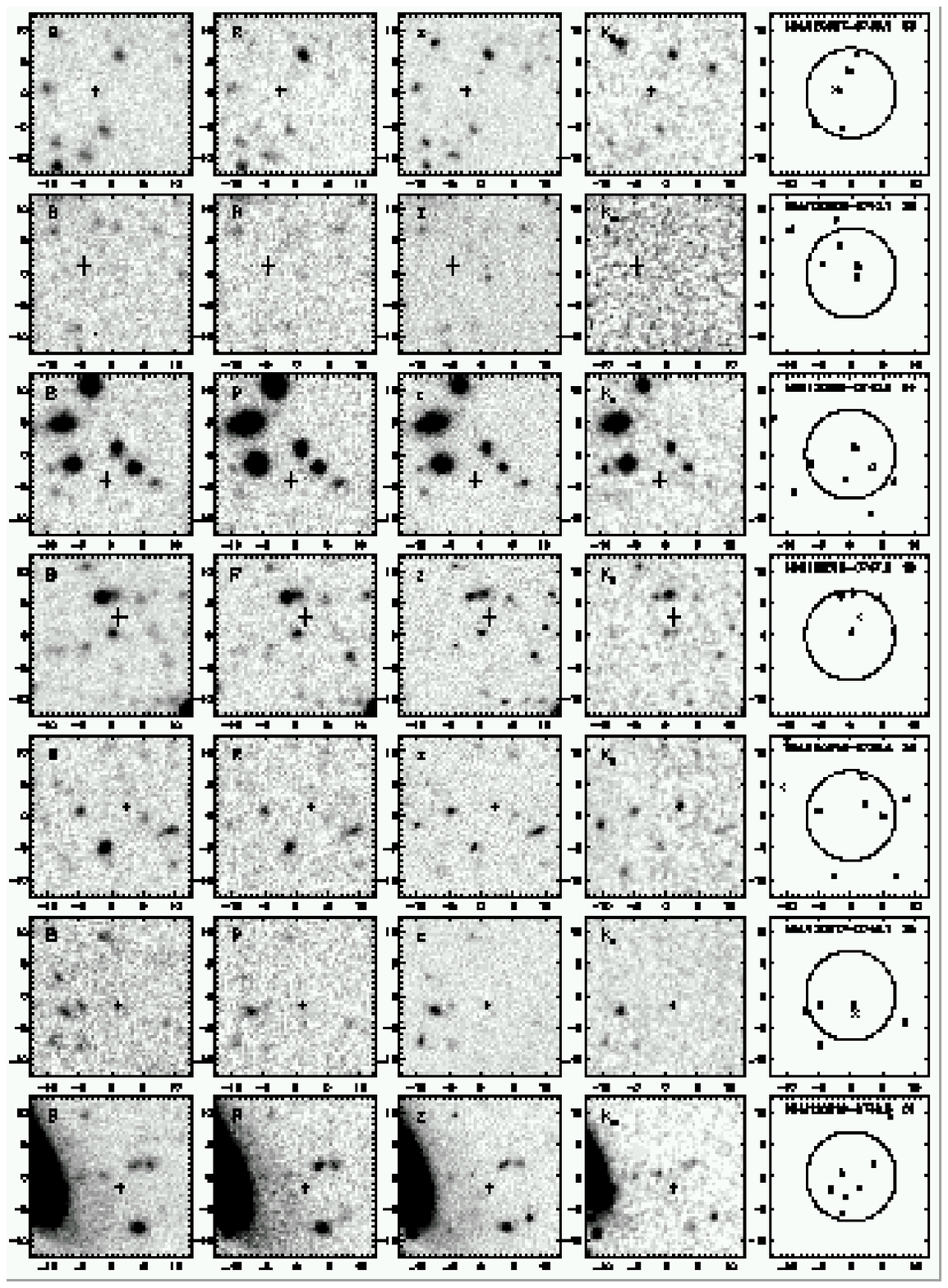}
\newpage
\epsscale{1.0}
\plotone{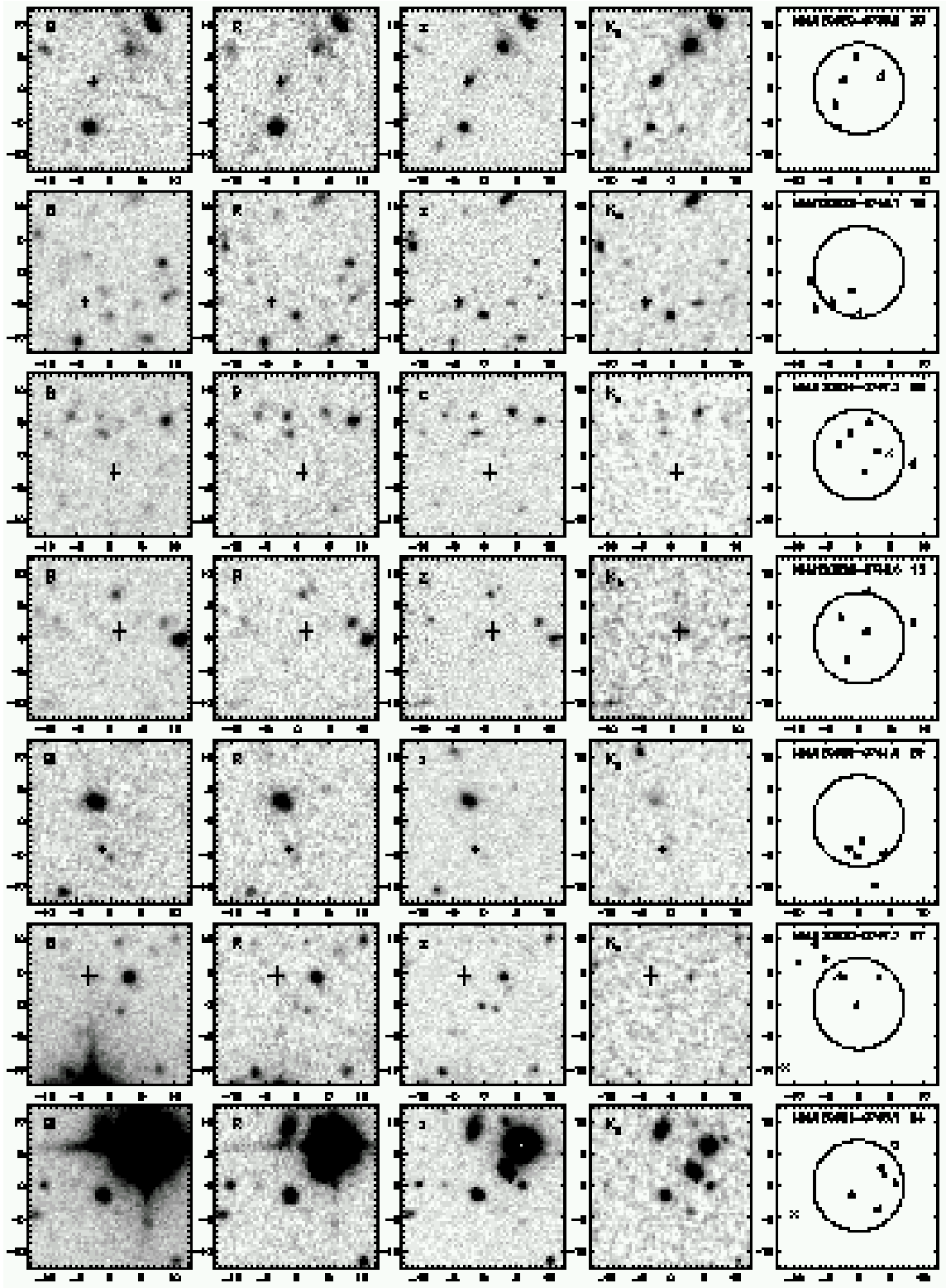}
\newpage
\epsscale{1.0}
\plotone{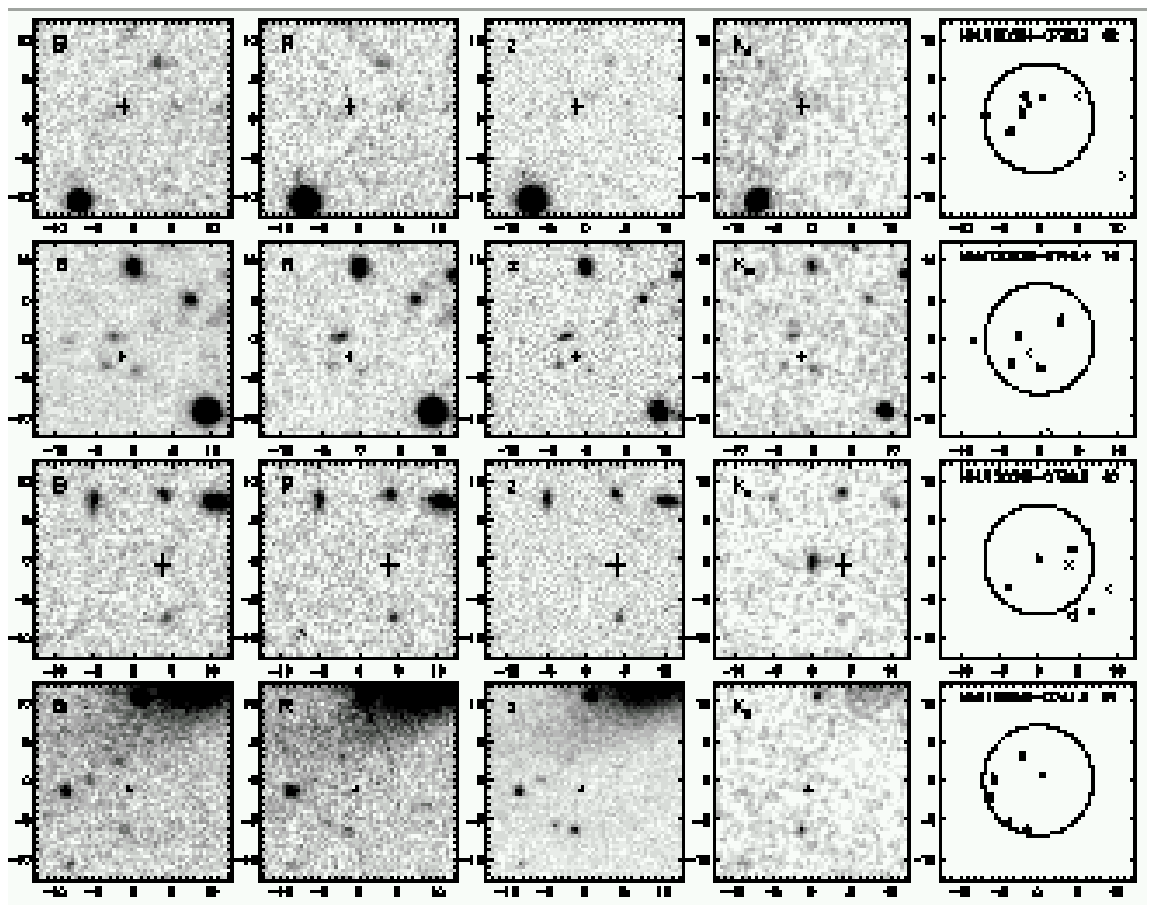}
\newpage
\epsscale{1.0}
\plotone{f3.eps}
\newpage
\epsscale{0.6}
\plotone{f4.eps}
\newpage
\epsscale{1.0}
\plotone{f5.eps}
\newpage
\epsscale{1.0}
\plotone{f6.eps}
\newpage
\epsscale{0.7}
\plotone{f7.eps}
\newpage
\epsscale{0.7}
\plotone{f8.eps}
\newpage
\epsscale{1.0}
\plotone{f9.eps}
\end{document}